\documentclass[reprint,superscriptaddress,aps,prb,twocolumn,floatfix]{revtex4-2} %
\usepackage{setspace}
\usepackage{amsmath}
\usepackage{bm}
\usepackage{graphicx}
\usepackage[nearskip,margin = 0pt]{subfig}
\usepackage{verbatim}
\usepackage{amsfonts}
\usepackage{amssymb}
\usepackage{textcomp}
\usepackage{mathrsfs}
\usepackage{mathtools}
\usepackage{url}
\usepackage{caption}
\usepackage{color}
\usepackage{upgreek}

\usepackage{xcolor}
\usepackage[colorlinks,linkcolor=blue,anchorcolor=blue,citecolor=blue,urlcolor=black]{hyperref}
\usepackage{ragged2e}
\usepackage{float}
\usepackage{lineno}
\DeclareGraphicsExtensions{.pdf,.eps,.png,.jpg,.mps}
\begin{document}

\title{Anti-resonant acoustic waveguides enabled tailorable Brillouin scattering on chip}
\author{Peng Lei$^{1}$, Mingyu Xu$^{1}$, Yunhui Bai$^{1}$, Zhangyuan Chen$^{1}$, and Xiaopeng Xie$^{1,\dagger}$ \\
\vspace{3pt}
$^1$State Key Laboratory of Advanced Optical Communication Systems and Networks,\\School of Electronics, Peking University, Beijing 100871, China\\
\vspace{3pt}
Corresponding authors: $^\dagger$xiaopeng.xie@pku.edu.cn.}



\maketitle
\noindent
\large\textbf{Abstract} \\
\normalsize\textbf{
Empowering independent control of optical and acoustic modes and enhancing the photon-phonon interaction, integrated photonics boosts the advancements of on-chip stimulated Brillouin scattering (SBS). However, achieving acoustic waveguides with low loss, tailorability, and easy fabrication remains a challenge. Here, inspired by the optical anti-resonance in hollow-core fibers, we propose suspended anti-resonant acoustic waveguides (SARAWs) with superior confinement and high selectivity of acoustic modes, supporting both forward and backward SBS on chip. Furthermore, this structure streamlines the design and fabrication processes. Leveraging the advantages of SARAWs, we have showcased a series of record-breaking results for SBS within a compact footprint on the silicon-on-insulator platform. For forward SBS, a centimeter-scale SARAW supports a large net gain exceeding 6.4 dB. For backward SBS, we have observed an unprecedented Brillouin frequency shift of 27.6 GHz and a mechanical quality factor of up to 1,960 in silicon waveguides. This paradigm of acoustic waveguide propels SBS into a new era, unlocking new opportunities in the fields of optomechanics, phononic circuits, and hybrid quantum systems.
}

\begin{figure*}[ht]
\centering
\captionsetup{singlelinecheck=no, justification = RaggedRight}
\hspace*{-0.7cm}
\includegraphics[width=17cm]{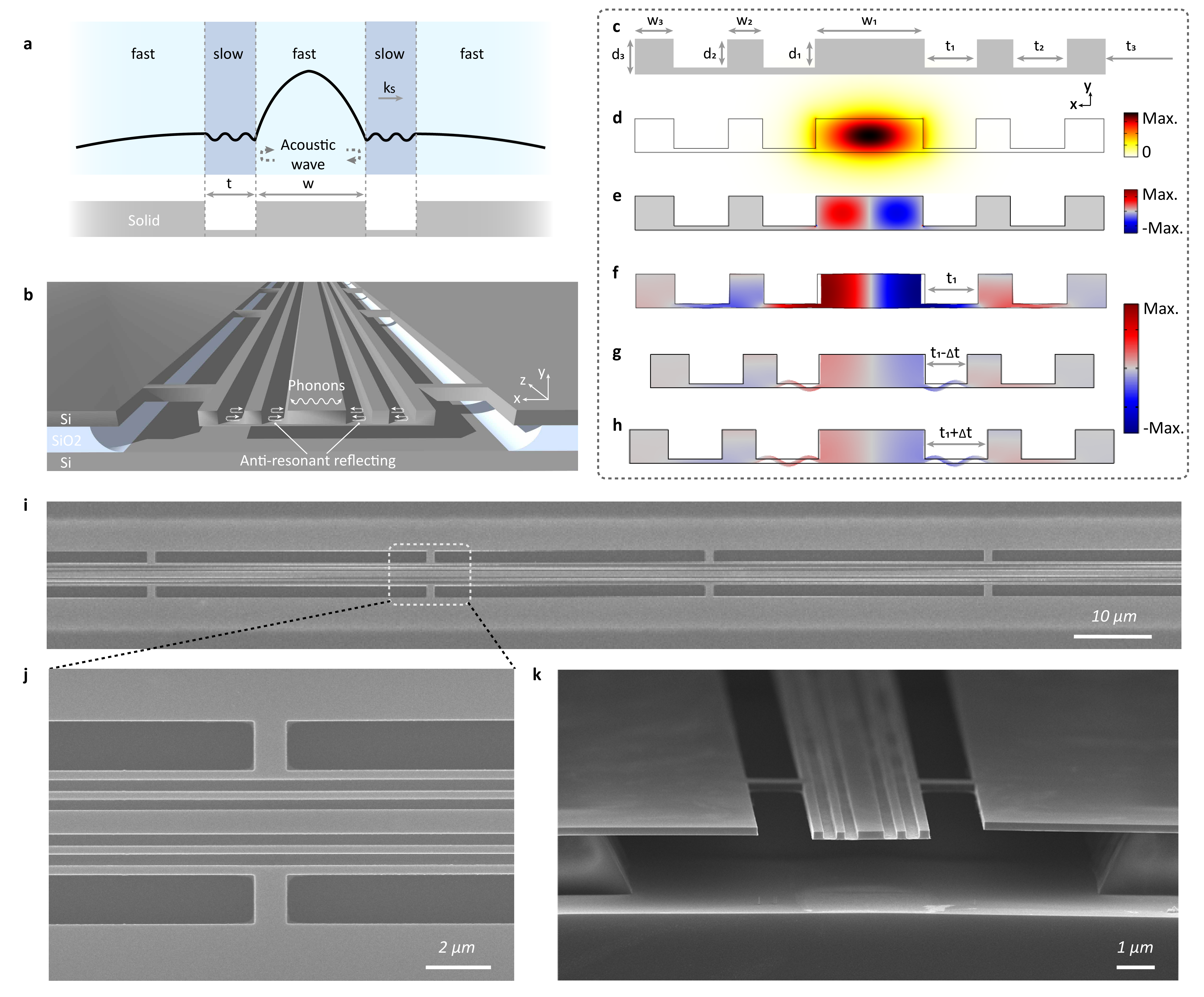}
\caption{\textbf{Suspended anti-resonant acoustic waveguide on the SOI platform. a} The principle of anti-resonant reflection and its implementation on chip. \textbf{b} Schematic of the SARAW. \textbf{c} Cross-section of the SARAW and its geometric parameters. These parameters are axisymmetric. W$_i$ (i = 1, 2, 3) corresponds to the width of the unetched region, t$_i$ corresponds to the thickness of the etched slot, and d$_i$ corresponds to the etching depth of the etched slot. Benefiting from the loading-effect etching process, wherein a thicker t$_i$ corresponds to a deeper d$_i$, we can control the etching depth d$_i$ by adjusting the slot thickness t$_i$. t$_3$ is set to sufficiently large to achieve overetching, which enables the hydrofluoric acid solution to remove the silica substrate. Consequently, the waveguides are fully suspended in air. \textbf{d} The electric field of the optical mode. \textbf{e} The x component of the electrostrictive force. \textbf{f} The elastic displacement field of the optimized acoustic mode. Here, slot thickness t$_1$ meets the anti-resonance condition. Fig.\textbf{e-f} use the optimized parameters for the forward SBS with a waveguide width W$_1$ of 700 nm. The detailed geometric parameters are shown in Supplementary Section III. \textbf{g,h} The elastic displacement field when acoustic mode under the resonance condition. In Fig.\textbf{f-h}, the color bar corresponds to the magnitude of the x component of the elastic displacement field. The deformations depict the cross-sectional displacement fields. \textbf{i,j} Top-down SEM image of the fabricated SARAW (\textbf{i}) and a magnified view (\textbf{j}). \textbf{k} Cross-sectional SEM image of the fabricated SARAW. Scale bars, 10 $\upmu$m (\textbf{i}); 2 $\upmu$m (\textbf{j}); and 1 $\upmu$m (\textbf{k}).
}
\label{fig1}
\end{figure*}

\vspace{3pt}
\maketitle
\noindent
\large\textbf{Introduction} \\
\normalsize
\noindent Brillouin nonlinearities arise from the coupling of photons and phonons \cite{Review1,Review2,Review3,SIM}. Due to its unique characteristics derived from acoustic phonons, SBS has found widespread applications in the realms of signal generation and processing \cite{Filter2,IMP2,IMP3,IMP5,Delay2,Delay3,IMP4,IMP1,Sensor1,Sensor2,Sensor3,Laser4,IOSP3,IOSP2}, such as narrow linewidth laser \cite{Laser4,IOSP3}, microwave photonic filters \cite{Filter2,IMP2,IMP3,IMP5}, slow and fast light \cite{Delay2,Delay3,IMP4}, distributed sensing \cite{Sensor1,Sensor2,Sensor3}, and optical non-reciprocal devices \cite{IOSP2}. Recently, integrated photonics has enabled better confinement and manipulation of optical and acoustic modes \cite{Mat1,Mat2,Mat3,Mat4,Mat5,Mat6,Mat7,Mat8}, ushering in a new era of photon-phonon interaction. 

While optical mode confinement methods are well established, how to confine acoustic modes is still evolving, which now primarily relies on three strategies. The first approach is acoustic total internal reflection, which is generally applied in materials with low stiffness \cite{Mat1,Mat2}, such as chalcogenides. Nonetheless, these materials are difficult to integrate with standard silicon photonic circuits. The second solution, based on acoustic impedance, involves isolating waveguides from the substrate to effectively prevent acoustic leakage \cite{Mat3,Mat4}. Whereas, this approach has difficulty in flexibly selecting acoustic frequencies, and necessitates complex fabrication processes. The third strategy utilizes phononic bandgaps to obstruct the spread of acoustic waves \cite{PP1,PP2,PP3}. However, its complex structures pose difficulties for design and manufacturing and result in low robustness.

Emerging anti-resonant reflection offers an alternative approach to achieve field confinement \cite{ARF1,ARF2,ARF3,ARRAW}. This approach has been successfully applied in hollow-core fibers \cite{ARF1,ARF2,ARF3}, effectively confining optical modes within the lower-refractive-index air core. Compared with photonic bandgap hollow-core fibers \cite{ARF1}, anti-resonant hollow-core fibers have a simpler structure and lower loss. Anti-resonant reflection has been theoretically proposed to confine acoustic modes in cylindrical waveguides \cite{ARRAW}. However, they are not compatible with integrated photonic platforms.

In this paper, we transfer the concept of anti-resonant reflection to integrated platforms and demonstrate a novel suspended anti-resonant acoustic waveguide (SARAW), proving its effectiveness in the confinement of phonons. By designing the anti-resonant structures, we can flexibly control the acoustic modes without affecting optical modes. This characteristic enables us to manipulate the acoustic mode distribution, select the eigenfrequency, and optimize the mechanical quality factor $Q_m$. Consequently, SARAWs can support both forward and backward SBS and have achieved a series of record-breaking results of SBS on the silicon-on-insulator (SOI) platform. For forward SBS, we have attained a record-high Brillouin gain coefficient $G_B$ of 3,530 W$^{-1}$m$^{-1}$, over 2,000 times larger than that of standard single-mode fibers. The threshold for achieving Brillouin net gain is remarkably low ($<$ 5 mW). Under a modest pump power, a large net gain of up to 6.4 dB can be realized with a compact footprint. These results highlight that SARAWs can significantly enhance the coupling strength of optical and acoustic waves. For backward SBS, we have observed a record-high Brillouin frequency shift of 27.6 GHz and an unprecedented $Q_m$ of 1,960 in integrated silicon waveguides. These breakthroughs show that acoustic waves in SARAWs can operate at frequencies extending to millimeter-wave bands, and exhibit extremely low propagation loss. In terms of fabrication and design, SARAWs eliminate the need for overlay exposure in the fabrication processes, thus suppressing the harmful inhomogeneous broadening of Brillouin resonance. It also allows for a simplified waveguide design process utilizing genetic algorithms \cite{GA1,GA2}.
 
\vspace{6pt}
\noindent
\large\textbf{Results}\\
\noindent
\normalsize
\textbf{Design and fabrication of SARAWs}\\
\noindent
We first introduce the principle of acoustic anti-resonant reflection with the planar waveguide, as depicted at the top of Fig.\ref{fig1}a. To confine the acoustic field within the middle layer, the side layers with slower acoustic velocities are incorporated as anti-resonant reflecting layers, which can be conceptualized as Fabry–Pérot cavities. By manipulating cavities thicknesses, the anti-resonance states can be achieved. Consequently, the acoustic wave is unable to traverse the side layers and is reflected to the middle layer, leading to the acoustic field distribution outlined by the black line in Fig.\ref{fig1}a. Generally, different velocity layers are composed of distinct materials \cite{ARF1,ARRAW}, such as silicon and silica. However, implementing this transverse multilayer structure on the chip is challenging. Here, we propose a new approach to address this issue. Considering the situation in the suspended solid membrane, as illustrated at the bottom of Fig.\ref{fig1}a, slots with thickness t are etched onto both sides of the central waveguide. In the slot regions, the boundary between air and solid geometrically softens the structural response of solid film \cite{Review3}, thus lowering its effective acoustic velocity and achieving anti-resonant reflecting layers. 

Following this approach, we propose the suspended anti-resonant acoustic waveguide as shown in Fig.\ref{fig1}b,c. The device consists of a central waveguide flanked by two sets of etched slots, and the entire suspended structure is supported by a series of tethers. Since silicon exhibits excellent optical and acoustic properties, such as a high refractive index and a low acoustic loss, and is compatible with complementary metal-oxide-semiconductor (CMOS) technology, it becomes a conducive medium for robust photon-phonon interactions. We deploy SARAWs on the SOI platform and demonstrate its improvement and tailorability on Brillouin nonlinearities.

We first investigate the properties of the SARAWs using forward SBS, which has a larger gain coefficient $G_B$ than backward SBS. Through the meticulous adjustment of anti-resonant structures using a genetic algorithm \cite{GA1,GA2} (see Methods), SARAWs can be optimized to achieve the largest $G_B$. We showcase the SARAW with a central waveguide width W$_1$ of 700 nm, see Supplementary Section III for detailed parameters. The SARAW supports a fundamental transverse electric-like optical mode (Fig.\ref{fig1}d), and the corresponding electrostrictive force is depicted in Fig.\ref{fig1}e. The acoustic mode is squeezed towards the central waveguide, as shown in Fig.\ref{fig1}f, indicating the effectiveness of anti-resonant reflection in acoustic mode confinement. Furthermore, the similar distribution of the electrostrictive force and acoustic mode suggests a large photon-phonon overlap, closing to that of the fully suspended rectangle waveguide. To further demonstrate the effect of anti-resonant reflection, we introduce a variation (t/t$_1$ = 20 $\%$) to the slot thickness (t$_{1}$), making SARAWs to operate in resonance conditions. As shown in Fig.\ref{fig1}g,h, the acoustic intensity at the central waveguide decreases sharply, and a significant portion of the acoustic energy leakages into the slot region, inducing a significant increase in the deformation. This results in a reduction in photon-phonon overlap. Hence, by adjusting the geometric parameters, SARAWs can be flexibly switched between the anti-resonance and resonance states, effectively manipulating the acoustic confinement and the photon-phonon coupling strength. Moreover, this manipulation does not influence the optical mode and can be performed independently.

SARAWs also streamline the fabrication processes and play a significant role in reducing the inhomogeneous broadening of Brillouin resonance. Previous suspended SBS waveguides typically necessitate overlay exposure \cite{Mat4,PP2}, involving at least two exposure steps and two etching steps, and all steps call for exceptional alignment precision. Furthermore, the alignment mismatches in the overlay exposure process would introduce inhomogeneous broadening of Brillouin resonance, exacerbating SBS performance. Particularly, the degradation of alignment mismatches is significantly magnified in spiral waveguides (see Supplementary Section VIII). Here, we innovatively propose an etching technique based on the loading effect to fabricate the whole structure via a single exposure and etching step (see Methods and Supplementary Section IV). The loading effect \cite{LE1}, where the etch rate depends on pattern width in etching processes, generally acts as a barrier to achieving depth uniformity. Yet, this property provides the opportunity to attain different etching depths in a single etching step. Given the distinctive structure of the SARAW, which requires different slots depths (d$_{1}$ $\sim$ d$_3$), the loading-effect etching technique can be effectively applied, avoiding the costs and yield losses caused by multiple fabrication cycles. By excluding the overlay exposure requirement, the loading-effect etching technique can accommodate high fabrication precision ($<$ 5 nm), and mitigate the inhomogeneous broadening. By introducing rectangle spiral waveguides (see Supplementary Section VIII), we can achieve centimeter-scale-long waveguides in a compact footprint with improved consistency, yielding higher $Q_m$ and $G_B$ compared to straight waveguides. On top of these design and fabrication processes, the SARAWs are analyzed by scanning electron microscopy (SEM) (Fig.\ref{fig1}i-k), which demonstrate excellent structural stability and uniformity.

\begin{figure*}[ht]
\centering
\captionsetup{singlelinecheck=no, justification = RaggedRight}
\includegraphics[width=14cm]{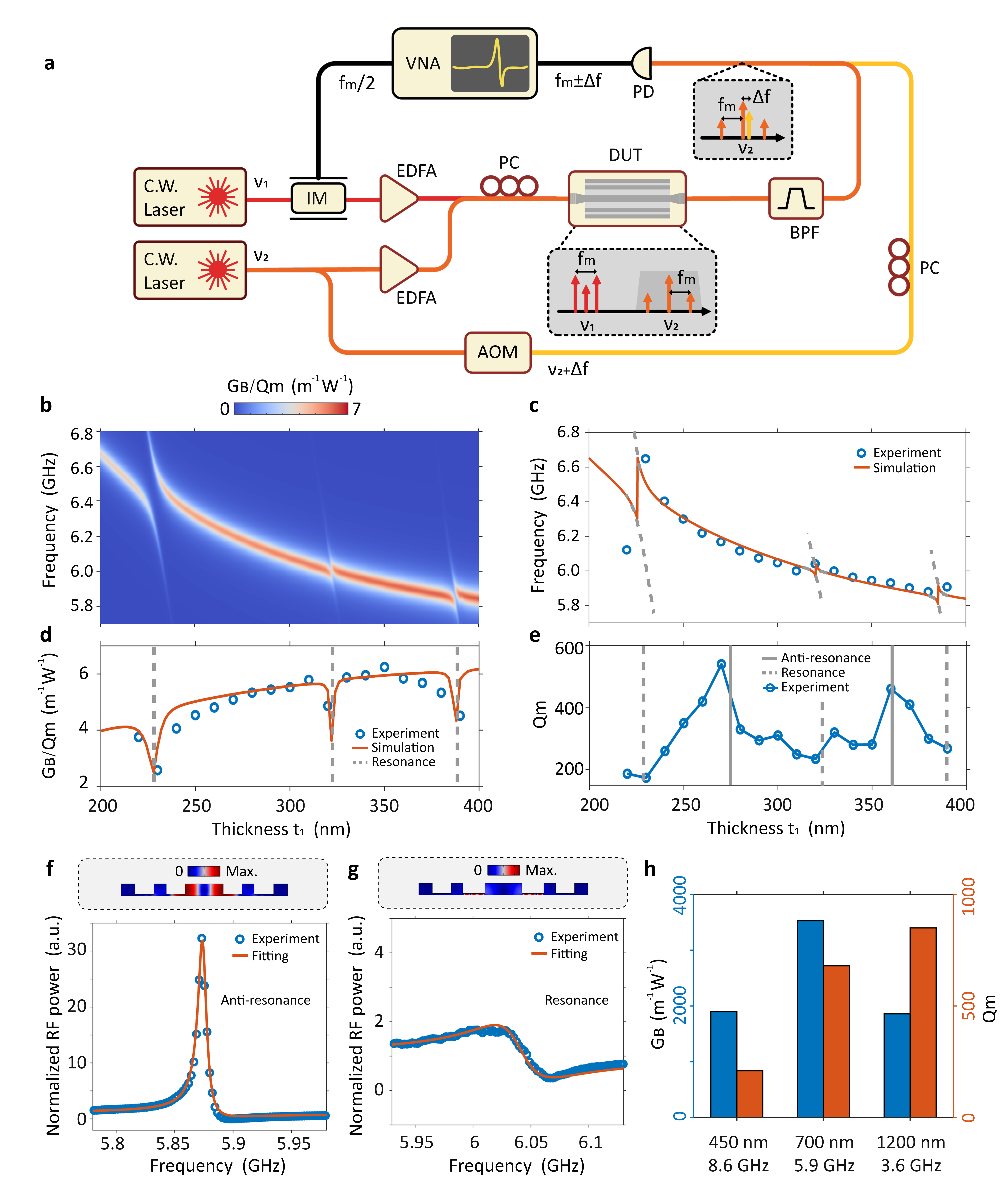}
\caption{\textbf{The demonstration of anti-resonance and experimental results. a} Diagram of the heterodyne FWM experiment. Two continuous-wave (C.W.) lasers operate at frequencies $\nu_1$ and $\nu_2$ (corresponding to wavelengths of 1550 and 1552 nm). The upper laser generates sidebands ($\nu_1\pm f_m/2$) via an intensity modulator (IM) operating at a carrier suppression point, with the modulation signal ($f_m/2$) provided by a vector network analyzer (VNA). After amplified by an erbium-doped fiber amplifier (EDFA), these sidebands serve as the pump light for the device under test (DUT). The second C.W. laser is split into two paths by a coupler. The upper branch is amplified by an EDFA as the probe light. The lower branch undergoes a frequency shift of $\Delta f$ via an acousto-optic modulator (AOM) as a reference signal. Within the DUT, the pump light ($\nu_1\pm f_m/2$) and the probe light ($\nu_2$) generate two sidebands at $\nu_2\pm f_m$ through FWM (the gray area beneath the DUT). To measure the intensity variations of the FWM-generated sidebands when $f_m$ scans over the Brillouin frequency shift, $\nu_2\pm f_m$ are filtered out via a cavity band-pass filter (BPF) (the dark gray trapezoidal box on the right below the DUT), and then mixed with $\nu_2+\Delta f$ on a photodetector (PD), yielding signals at $f_m\pm\Delta f$. The corresponding FWM variation curves can be obtained on the VNA, corresponding to Stokes ($f_m+\Delta f$) and anti-Stokes ($f_m-\Delta f$) components. \textbf{b} The influence of resonance/anti-resonance condition on Brillouin frequency shift and coupling factor $G_B/Q_m$ with different slot thickness ($t_1$). To facilitate observation, the linewidth of the acoustic modes is set to 50 kHz. \textbf{c-e} Experimental and simulated results show the impact of anti-resonant reflection on Brillouin frequency shift, $G_B/Q_m$, and $Q_m$, respectively. \textbf{f-g} The elastic displacement magnitude (top) and measurement results (bottom) of the heterodyne FWM experiment (anti-Stokes sideband, $\nu_2+f_m$) in SARAWs under the anti-resonance condition (\textbf{f}) and the resonance condition (\textbf{g}). Here, the waveguide width W$_1$ = 700 nm. \textbf{h} The experimental attained $G_B$ and $Q_m$ of optimized SARAWs for W$_1$ = 450, 700, and 1,200 nm, with the acoustic eigenfrequency of 8.6, 5.9, and 3.6 GHz, respectively.
}
\label{fig2}
\end{figure*}

\vspace{3pt}
\noindent 
\maketitle
\textbf{Demonstration of anti-resonance with SARAWs}\\
\noindent Initially, we validated the tailorability of SARAWs through simulation in terms of the acoustic frequency and mode distribution. For the anti-resonant structure, the slot thickness t$_1$ can effectively control the anti-resonance state and influence the acousto-optic interaction strength. Therefore, we varied the slot thickness t$_1$ of the SARAW from 200 to 400 nm, with the other parameters kept the same as in Fig.\ref{fig1}f. Through the simulation (see Methods and Supplementary Section III), we tracked the acoustic frequency and coupling factor $G_B/Q_m$, which reflects the photon–phonon overlap. The simulation results, shown in Fig.\ref{fig2}b, indicate a decrease in the frequency of the Brillouin-active acoustic mode from 6.8 GHz to 5.9 GHz as the thickness t$_1$ increases. This trend results from the influence of another acoustic mode (see Supplementary Section III). During the sweep, three resonance conditions are met (t$_1$ = 225, 325, and 338 nm), with two anti-resonance conditions in between. When slot thickness t$_1$ meets the resonance condition, avoided mode crossing \cite{AvoidedMC} is observed. This phenomenon arises from the strong coupling between the acoustic mode in the central waveguide and the shear modes in the slot regions, leading to a significant decrease in photon-phonon overlap. The uneven appearance of resonance conditions when t$_1$ increases is mainly attributed to the consideration of the loading effect in the simulations (see Supplementary Section III).

\begin{figure}[b]
\centering
\captionsetup{singlelinecheck=no, justification = RaggedRight}
\hspace*{-0.7cm}
\includegraphics[width=8.0cm]{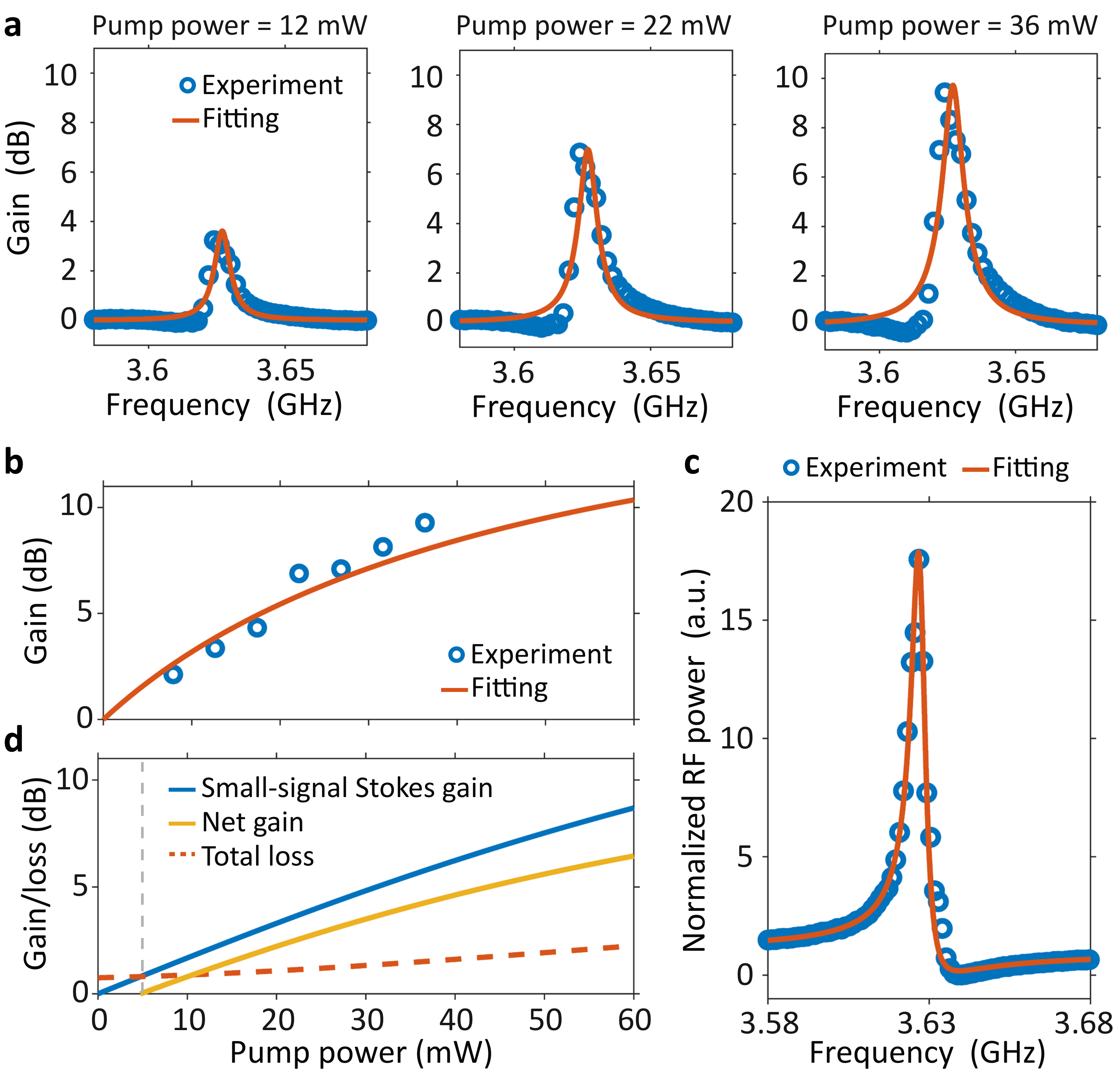}
\caption{\textbf{Measurement of Brillouin gain. a} Normalized Stokes transmitted spectra for pump powers of 12, 22, and 36 mW. \textbf{b} The experimental and simulation results of three-tone direct gain. \textbf{c} The heterodyne FWM experiment result of the same SARAW. \textbf{d} Small-signal Stokes gain and net gain. The threshold for net gain is 5 mW.
}
\label{fig3}
\end{figure}

Using the parameters in the simulation, we fabricated a series of 5-mm-long SARAWs and conducted the heterodyne four-wave mixing (FWM) experiment \cite{Mat4,PP2}. This experimental setup, shown in Fig.\ref{fig2}a, provides excellent sensitivity and avoids the impact of coupling loss fluctuations, thus ensuring robust and reliable results. We extract Brillouin frequency shift, $G_B$, and $Q_m$ by fitting the obtained Fano-like spectral lines (see Supplementary Section V). In order to clearly compare the simulation and experimental results, acoustic frequency and coupling factor $G_B/Q_m$ are depicted in Fig.\ref{fig2}c,d respectively. Avoided mode crossing and decrease of coupling factor $G_B/Q_m$ under resonance conditions are observed, showing good agreement with the simulation. As shown in Fig.\ref{fig2}e, the dips and peaks of mechanical quality factor $Q_m$ illustrate the noticeable difference between resonance and anti-resonance conditions for the acoustic confinement effect. The difference is further illustrated by Brillouin spectra obtained by the heterodyne FWM experiment under corresponding conditions (Fig.\ref{fig2}f,g). Under the anti-resonance condition (Fig.\ref{fig2}f), the fitted $G_B$ and $Q_m$ reach up to \mbox{3530 W$^{-1}$m$^{-1}$} and 680, indicating that the acoustic mode is effectively confined within the central waveguide. While under the resonance condition (Fig.\ref{fig2}g), both $G_B$ and $Q_m$ decrease dramatically by over 80\%, reaching 540 W$^{-1}$m$^{-1}$ and 120, respectively, as the acoustic mode energy disperses into the slot regions. The above simulation and experimental results solidly validate the effectiveness of SARAWs for acoustic confinement and manipulation. 

Utilizing the manipulating ability and frequency selectivity of anti-resonance, we optimize and fabricate 5-mm-long SARAWs with different central waveguide widths (W$_1$ = 450, 700, and 1,200 nm) to obtain the respective maximum $G_B$. The experimentally attained Brillouin frequency shift, $G_B$, and $Q_m$ are shown in Fig.\ref{fig2}h. As the waveguide width increases, the waveguide acquires a larger mode field area and becomes less sensitive to sidewall roughness. Therefore, the eigenfrequency and coupling factor $G_B/Q_m$ decrease gradually, while $Q_m$ increases accordingly. The 450-nm SARAW exhibits the highest $G_B/Q_m$ of 8.57, but suffers more from the inhomogeneous broadening caused by sidewall roughness, resulting in a low $Q_m$. The 700-nm SARAW achieve a record-breaking $G_B$ of 3530 W$^{-1}$m$^{-1}$ in silicon waveguides with a balance of $G_B/Q_m$ and $Q_m$. And the 1,200-nm SARAW shows a higher $Q_m$ of up to 850. Consequently, SARAWs can be flexibly extended to different waveguide widths, allowing for the attainment of distinct acoustic frequencies, $G_B$ and $Q_m$.

\vspace{3pt}
\noindent 
\maketitle
\textbf{Large Brillouin net gain in SARAWs}\\
\noindent The Brillouin net gain, which takes the Brillouin gain coefficient $G_B$ and optical loss into account, is typically a comprehensive figure of merit for Brillouin-based applications. Hence, we conducted further design and experiment to illustrate the potential of SARAWs in achieving large Brillouin net gain. Benefiting from the properties of low optical loss and minimal inhomogeneous broadening, SARAWs with a central waveguide width of 1,200 nm are suitable for attaining large net gain. We examine Brillouin interactions in a 2.5-cm-long rectangular spiral SARAW with a compact footprint of \mbox{1,900 $\upmu$m $\times$ 460 $\upmu$m}. 

We carried out a three-tone direct gain experiment \cite{PP2} (see Supplementary Section VI) to obtain the net gain, owing to its simpler setup compared to traditional small-signal gain experiment. Fig.\ref{fig3}a shows three Brillouin gain spectra measured for pump powers of 12, 22, and 36 mW. These gain spectra, which illustrate the relative change in the probe intensity, reveal a Brillouin resonance with a high mechanical quality factor ($Q_m$ = 660, see Supplementary Section VI for a detailed fitting process) at 3.63 GHz. The probe intensity is significantly amplified by approximately 10 dB under 36 mW pump power. There is a slight discrepancy between the experimental results and the fittings, which could be attributed to the phase drift of the modulator and the acoustic eigenfrequency drift. Fig.\ref{fig3}b shows the Stokes gain at Brillouin resonance as a function of pump power. To satisfy the small signal approximation, the pump power is kept below 40 mW. By numerically solving the three-tone coupling equations (see Supplementary Section VI), the fitted $G_B$ is determined as 1,700 W$^{-1}$m$^{-1}$ ( Fig.\ref{fig3}b). It is of great consistency with the independent measurement of the same waveguide performed through heterodyne FWM measurement ($G_B$ = 1,670 W$^{-1}$m$^{-1}$, $Q_m$ = 650), as is shown in Fig.\ref{fig3}c. Compared to the 5-mm-long SARAW with the same structure parameters (Fig.\ref{fig2}h, central waveguide width W$_1$ = 1200 nm), Brillouin gain coefficient $G_B$ and mechanical quality factor $Q_m$ here are slightly reduced due to the inhomogeneous broadening under long waveguide length. Based on the results from the above experiments (Fig.\ref{fig3}b,c), the small-signal Stokes gain and the net gain are numerically calculated (see Supplementary Section VI) and shown in Fig.\ref{fig3}d. Benefiting from the low optical loss (0.3 dB/cm, see Supplementary Section I) and large Brillouin gain, the threshold for net gain is below 5 mW. Considering the maximum on-chip pump power of 57 mW constrained by EDFA and coupling loss (single-sideband 5.4 dB), SARAWs can support a net gain of up to 6.4 dB and an on-off gain of 8.7 dB. These results mark the present apex of Brillouin gain levels in silicon waveguides and can be further improved with higher on-chip power and longer waveguide length. 

\begin{figure}[b]
\centering
\captionsetup{singlelinecheck=no, justification = RaggedRight}
\hspace*{-0.7cm}
\includegraphics[width=8.0cm]{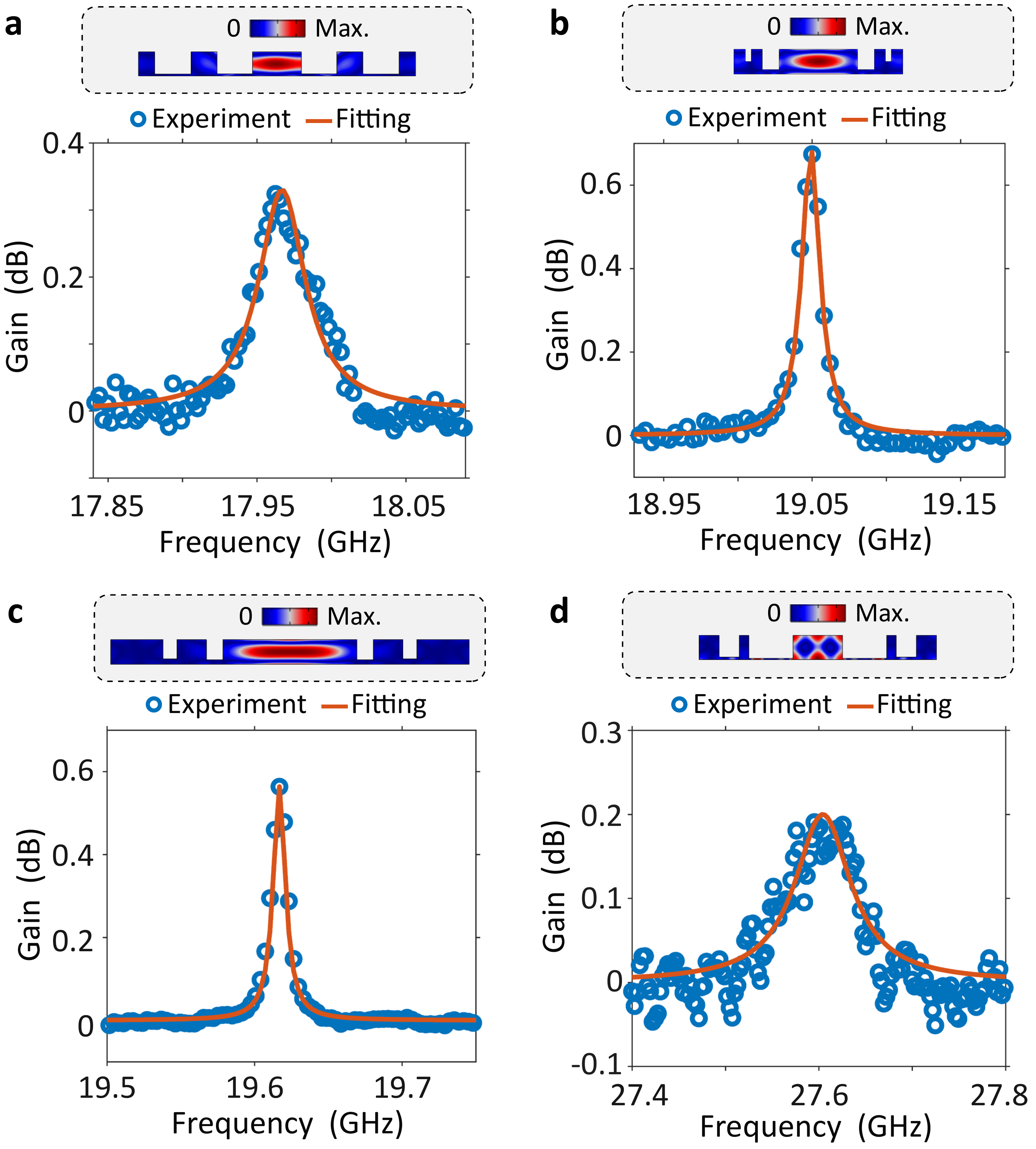}
\caption{\textbf{Measurement of Backward SBS. a-d} The elastic displacement magnitude of the corresponding acoustic mode field (upper gray region) and backward SBS gain spectra (lower). Figures \textbf{a-c} correspond to the optimized acoustic modes with central waveguide width W$_1$ = 450, 700, and 1200 nm, using the genetic algorithm. Figure \textbf{d} illustrates the outcomes obtained through optimization for higher-order acoustic modes within a 450-nm SARAW. The values of Brillouin gain coefficient $G_B$ (W$^{-1}$m$^{-1}$) and mechanical quality factor $Q_m$ are: (530, 470); (600, 1,300); (430, 1,960); (480, 380), corresponding to figure \textbf{a-d}, respectively.
}
\label{fig4}
\end{figure}

\vspace{3pt}
\noindent 
\maketitle
\textbf{Backward SBS in SARAWs}\\
\noindent Backward SBS, mediated by the interaction of two counter-propagating optical waves with a phase-matched longitudinal acoustic wave, exhibits distinct characteristics compared to forward SBS. Acoustic modes in backward SBS exhibit higher acoustic frequency and quality factor. The longitudinal wave nature of the backward SBS acoustic mode necessitates an extensive waveguide length to build up to its full strength. However, existing suspended waveguides inevitably introduce periodic supporting structures \cite{Mat4,Mat10}, leading to the dissipation of acoustic waves. Meanwhile, the material property of the photoelastic coefficient makes it challenging to observe backward SBS in silicon waveguides \cite{Review2}. The maximum on-off gain, $G_B$, and Brillouin frequency shift for backward SBS in silicon waveguides achieved so far are only 0.2 dB, 359 W$^{-1}$m$^{-1}$, and 13.7 GHz \cite{Mat3}, respectively. However, in SARAWs, the anti-resonant structures separate the acoustic wave from the periodic supporting tethers. This separation enables the acoustic wave to build up to its full strength, thereby facilitating robust support for backward SBS.

We have designed and fabricated 2-cm-long SARAWs with central waveguide widths W$_1$ = 450, 700, and 1,200 nm. Utilizing the setup in Supplementary Section VII, we have observed a bell-shaped acoustic mode under all three waveguide widths (Fig.\ref{fig4}a-c). Same as the trends in forward SBS, as the central waveguide width W$_1$ increases, the photon-phonon overlap decreases and the mechanical quality factor increases. With a balance of $G_B/Q_m$ and $Q_m$, the 700 nm SARAW achieves a backward Brillouin gain coefficient $G_B$ of 600 W$^{-1}$m$^{-1}$ (Fig.\ref{fig4}b), nearly doubling the current record observed in silicon waveguides \cite{Mat3}. And surprisingly, the 1,200-nm SARAW (Fig.\ref{fig4}c) attains an unprecedented $Q_m$ of up to 1,960 in integrated waveguides, significantly exceeding that of forward SBS acoustic modes. Furthermore, a record-high acoustic frequency of 27.6 GHz is observed in another acoustic mode (Fig.\ref{fig4}d) utilizing a 450-nm SARAW. The geometric parameters used here have been meticulously optimized for this specific acoustic mode, distinguishing from the parameters used in Fig.\ref{fig4}a. It underscores the selectivity of SARAWs for specific acoustic modes and frequencies. However, this acoustic mode exhibits lower $G_B$ of 480 W$^{-1}$m$^{-1}$ and $Q_m$ of 380. It may be attributed to the fact that a significant portion of the acoustic mode is distributed near the waveguide boundaries, which results in a smaller acousto-optic overlap and makes it more vulnerable to surface roughness. Thus this mode has not been observed in SARAWs with central waveguide widths W$_1$ of 700 and 1,200 nm.

The record-breaking results above demonstrate that SARAWs can flexibly enhance and manipulate acoustic modes for backward SBS. As a result, SARAWs support acoustic modes with high group velocity ($>$ 3,000 m/s), and the corresponding acoustic decay length of over 50 $\upmu$m, offering a low-loss and long-distance transmission platform for acoustic waves.

\vspace{6pt}
\noindent
\large\textbf{Discussion} \\
\normalsize
\noindent
Leveraging genetic-algorithm-based optimization scheme and loading-effect-based fabrication technique, SARAWs can be rapidly designed and fabricated. For design scheme, the genetic algorithm eliminates the need for time-consuming parameter sweep and greatly simplifies the SARAW design process. Moreover, it allows for the rapid and specific optimization of various merit figures in SBS. This versatility is crucial for SBS-based applications with different requirements depending on the use-case. For instance, SBS-based microwave photonic filters \cite{Filter2,IMP2,IMP3,IMP5} with different center frequencies, bandwidths, and out-of-band suppression ratios demand diverse Brillouin frequency shift, mechanical quality factor $Q_m$, and Brillouin gain coefficient $G_B$. Meanwhile, the innovative loading-effect-based etching technique minimizes the manufacturing cost while maintaining excellent fabrication precision. It lessens the degradation of Brillouin resonance caused by inhomogeneous broadening while minimizing the overall footprint. Furthermore, this scheme is portable to other material platforms, rendering it highly suitable for integrated acousto-optic devices.

In our work, the net gain in SARAWs is mainly constrained by the coupling loss and the roughness of the waveguide sidewall, which are limited by our fabrication capabilities. These limitations hinder further increases in on-chip pump power and mechanical quality factor $Q_m$, and impede further reductions in optical propagation loss. However, referring to previous work \cite{Mat3}, the utilization of CMOS pilot line in the foundry could potentially yield lower coupling loss, higher $Q_m$, and lower optical loss. These advancements promise a performance enhancement of more than twofold for current net gain results.

We would like to emphasize that SARAWs exhibit exceptional characteristics in backward SBS. The backward SBS acoustic mode exhibits a significantly larger group velocity and an extended phonon lifetime than the forward SBS, resulting in a large acoustic decay length. This is also a crucial characteristic for the acoustic waveguides, making SARAWs a promising candidate for supporting the burgeoning field of phononic circuits \cite{PC1}. Furthermore, the promotions in acoustic mode eigenfrequency and $Q_m$ underscore the potential of SARAWs. A maximum $Q_m$ of 1,960 and an unprecedented acoustic eigenfrequency of 27.6 GHz are particularly advantageous for microwave photonics and signal processing devices for 5G and future communication systems, including microwave generations \cite{IMP1} and acoustic wave filters \cite{AWF1}. In addition, the acoustic frequency of 19.6 GHz and $Q_m$ of 1960 indicate a product of frequency and quality factor $f\times Q_m$ exceeding 3.8$\times$10$^{13}$ Hz, which is important for quantum systems. It is expected to be further improved in phonon lasers \cite{PC3} based on SARAWs.

In summary, SARAWs incorporate innovations in acoustic anti-resonance theory, intelligent algorithm empowered design, and nanofabrication technique, offering a comprehensive solution for integrated photon-phonon interaction platforms. It heralds a new era for SBS, paving the way for breakthroughs in optomechanics \cite{OptoM1}, phononic circuits \cite{PC1}, and hybrid quantum systems \cite{HQS2}.

\vspace{6pt}
\noindent \textbf{Methods}\\
\begin{footnotesize}
\noindent \textbf{Device fabrication.} 
The SARAW was fabricated from the 220 nm silicon layer of a SOI wafer with the \mbox{$\langle100\rangle$} crystal orientation utilizing electron-beam lithography (EBL) and inductively coupled plasma (ICP) etching. With loading-effect-based etching technique, SARAW can be fabricated with one exposure step and one etching step. Whereafter, the oxide undercladding was then released by immersion in \mbox{10 $\%$} hydrofluoric acid. The SARAW employs a rectangular spiral for long waveguides. The 2.5-cm-long waveguide has a footprint of 1,900 $\upmu$m $\times$ 460 $\upmu$m (see Supplementary Section IV).

\vspace{3pt}
\noindent 
\textbf{Genetic algorithm and finite element simulation.} First, we developed the main program and invoked the Genetic Algorithm Toolbox in MATLAB. Then, we adopted the simulation files from Ref.\cite{SIM}, and simulated acoustic and optical modes through the finite element solver COMSOL, calculating the total $G_B/Q_m$ produced by electrostrictive force and radiation pressure. Subsequently, the obtained $G_B/Q_m$ was further iterated by the genetic algorithm until the algorithm converged. It should be noted that due to the loading effect, slot thickness t$_i$ dictates the slot depth d$_i$ (Fig.\ref{fig1}c). Therefore, we measured the relationship curve between t$_i$ and d$_i$ caused by the loading effect through separate experiments, and then compensated it into the main program, ensuring that our optimization results were consistent with the actual fabricated waveguides (see Supplementary Section III).

\end{footnotesize}
\vspace{20pt}

\bibliography{ms.bib}


\end{document}



\title{Supplementary Information: Anti-resonant acoustic waveguides enabled \\tailorable Brillouin scattering on chip}
\author{Peng Lei$^{1}$, Mingyu Xu$^{1}$, Yunhui Bai$^{1}$, Zhangyuan Chen$^{1}$, and Xiaopeng Xie$^{1,\dagger}$ \\
\vspace{3pt}
$^1$State Key Laboratory of Advanced Optical Communication Systems and Networks,\\School of Electronics, Peking University, Beijing 100871, China\\
\vspace{3pt}
Corresponding authors: $^\dagger$xiaopeng.xie@pku.edu.cn.}
\maketitle



\date{\today}

\tableofcontents

\begin{figure*} [ht]
\centering
\captionsetup{singlelinecheck=no, justification = RaggedRight}
\includegraphics[width=12cm]{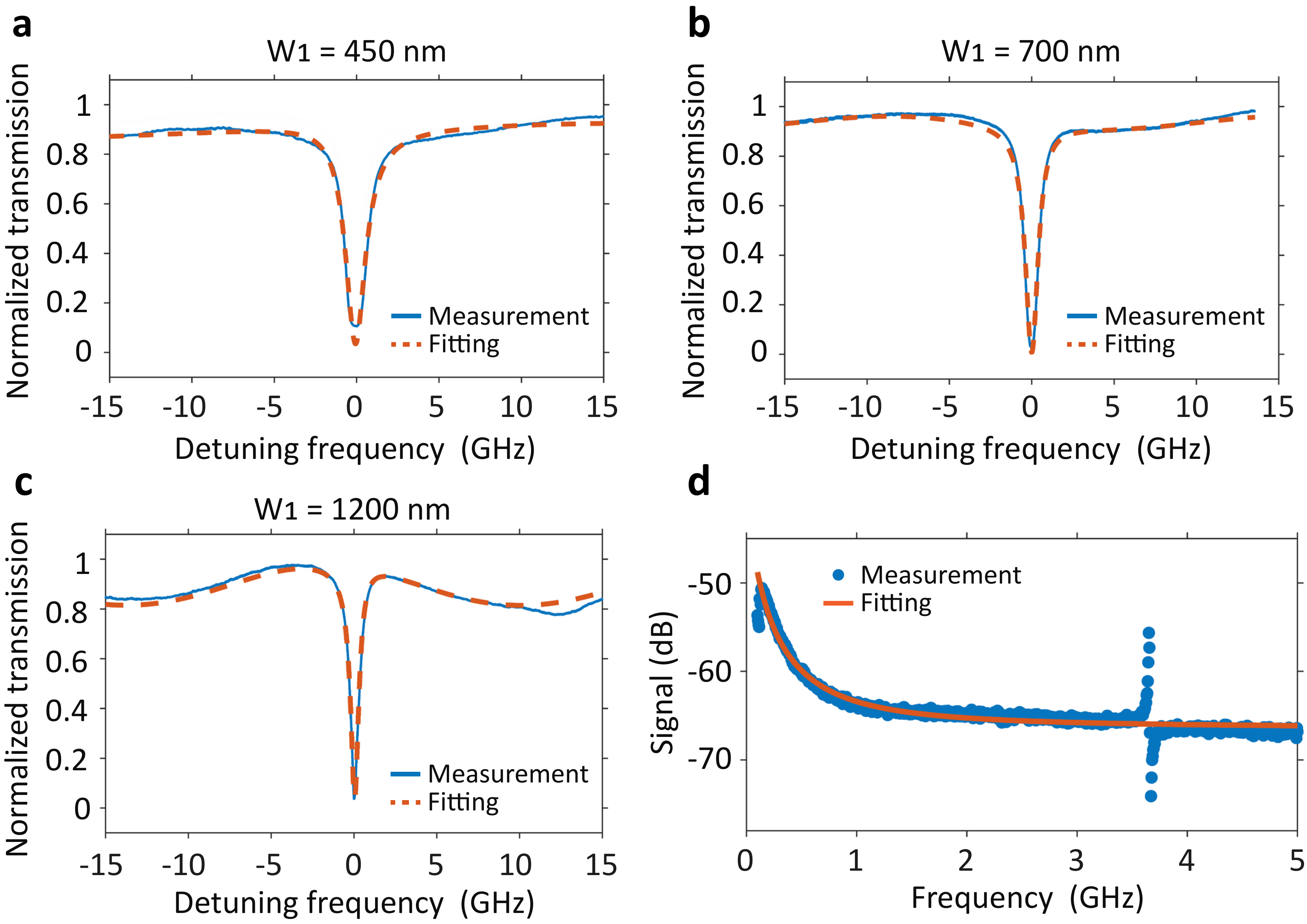}
\caption{\textbf{a-c} The optical quality factor fittings with FP background. \textbf{d} The measurement of free carrier lifetime using heterodyne FWM experiment. The Brillouin resonance is observed near 3.6 GHz, corresponding to the frequency of Brillouin resonance in Fig.3c in the main text.}
\label{FigS1}
\end{figure*}

\section{Linear and nonlinear loss model}
The optical linear and nonlinear losses are crucial parameters in the estimation of the Brillouin gain. The propagation loss of optical power ($P$) along z direction in silicon waveguides can be described by the following differential equation [1]
\begin{equation}
\label{propagateloss}
\frac{dP}{dz} = -\alpha P -\beta P^2-\gamma P^3.
\end{equation}
Here, $\alpha$ is the linear loss coefficient. $\beta$ and $\gamma$ are the nonlinear loss coefficients of two-photon absorption (TPA) and TPA-induced free-carrier absorption, respectively. 

The linear loss coefficient ($\alpha$) of the fundamental TE mode in SARAWs can be obtained by measuring the optical quality factor of micro-rings with the corresponding central waveguide width. This method is insensitive to fluctuations of coupling losses. Considering the Fabry-Pérot (FP) background caused by the grating coupler reflection, the transmission spectrum of the micro-ring resonator is given by [2, 3]:
\begin{equation}
\label{trans}
T_{total}(\Delta)=\lvert t_1 t_2\rvert^2 \frac{\lvert T_{res}(\Delta) \rvert^2}{\lvert 1-r_1 r_2 T_{res}(\Delta)^2 e^{-i \Delta/\omega_{FP}+i\phi} \rvert^2},
\end{equation}
where $t_1$ ($t_2$) and $r_1$ ($r_2$) are the transmission and reflection coefficients of the grating couplers, respectively. $T_{res}$ is the linear transmission function of the resonator. $\Delta$ is the detuning of the pump from the resonance frequency of the micro-rings. $\omega_{FP}$ is the free spectral range of the FP background, and $\phi$ is a phase offset. $T_{res}=1-\kappa_{ex}/(\kappa/2+i\Delta)$, with the total resonator loss $\kappa = \kappa_{in}+\kappa_{ex}$, the intrinsic loss rate $\kappa_{in}$, and the external coupling rate $\kappa_{ex}$. By fitting the quality factor of the fabricated micro-ring resonators with different central waveguide widths $W_1$ (Fig.2h in the main text) through Eq.\ref{trans}, we can obtain the intrinsic loss rate $\kappa_{in}$, and the derived intrinsic quality factors ($Q_{in}$) are 2.2$\times$10$^5$, 4.0$\times$10$^5$, and 8.5$\times$10$^5$ (FigS.\ref{FigS1}a-c), corresponding to $W_1$ = 450, 700, 1,200 nm separately. Utilizing the relation between $Q_{in}$ and $\alpha$ [4],
\begin{equation}
\label{linearloss}
Q_{in}=\frac{2\pi n_g}{\lambda\alpha},
\end{equation}
with the group index $n_g$ and the optical wavelength $\lambda$, the corresponding linear losses are 3.3, 1.6, and 0.3 dB/cm. A wider waveguide is less sensitive to sidewall roughness [5], resulting in lower $\alpha$. 

The nonlinear losses are mainly results from two-photon absorption and free-carrier absorption, and the nonlinear coefficients can be expressed as:
\begin{equation}
\label{nonlinearloss} 
\beta=\frac{\beta_{TPA}}{A_{eff}},\thickspace\gamma=\frac{\sigma\tau\beta_{TPA}}{2h\nu A_{eff}^2}, 
\end{equation}
with the bulk TPA coefficient $\beta_{TPA}$, the effective mode areas $A_{eff}$, the free-carrier absorption cross-section $\sigma$, the free-carrier lifetime $\tau$, and the energy of a single photon $h\nu$. Taking advantage of the lower linear loss characteristics of SARAW with a wider central waveguide width, we used SARAWs with a central waveguide width of 1,200 nm to achieve a large net gain and calculate their nonlinear loss coefficients. Through a full vectorial model [6], $A_{eff}$ is calculated to be 1.674$\times$10$^{-13}$ m$^2$. Referring to $\beta_{TPA}$ = 7.9$\times$10$^{-12}$ mW$^{-1}$ and $\sigma$ = 1.45$\times$10$^{-21}$ m$^2$ in bulk silicon [7], $\beta$ is 47 m$^{-1}$W$^{-1}$. By deploying the free-carrier lifetime measurement based on heterodyne four-wave mixing (FWM) experiment [1], the fitted $\tau$ is 2.6 ns (FigS.\ref{FigS1}d), and the calculated $\gamma$ is 4,150 m$^{-1}$W$^{-2}$. 

\section{Phase-matching and gain model of SBS}
\subsection{Phase-matching of SBS}
\begin{figure}[ht]
\centering
\captionsetup{singlelinecheck=no, justification = RaggedRight}
\hspace*{-0.2cm}
\includegraphics[width=9cm]{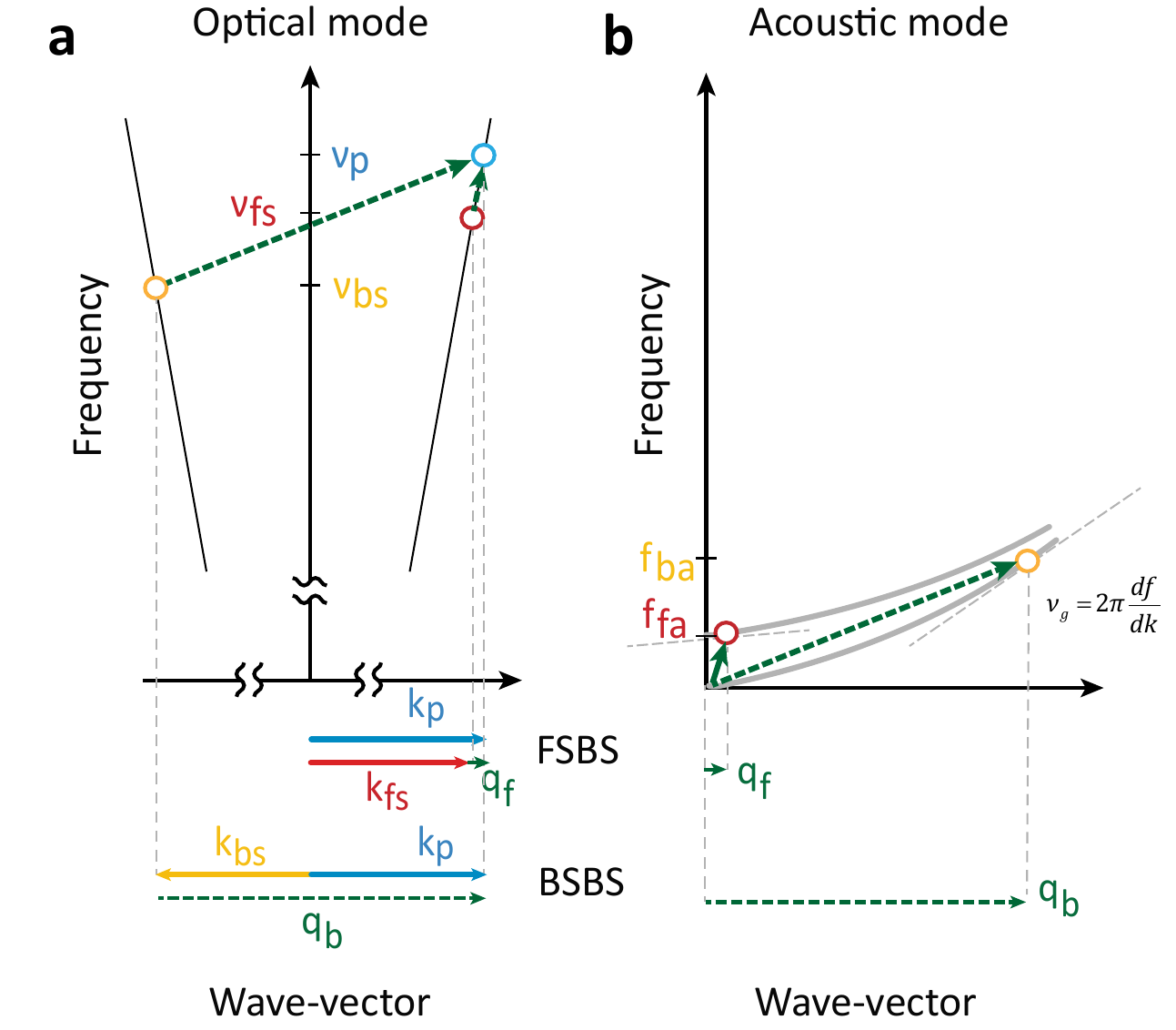}
\caption{\textbf{a} Energy conservation and phase-matching for forward and backward SBS. \textbf{b} Dispersion relations of acoustic modes for forward and backward SBS.
}
\label{FigS2}
\end{figure}

The process of SBS demands the conservation of energy and momentum, imposing strict phase-matching requirements among the pump, Stokes, and acoustic waves. Depending on the relative propagation directions of the pump and Stokes lights, intramodal SBS can be classified into forward SBS (FSBS) and backward SBS (BSBS), each corresponding to acoustic modes with distinct characteristics. The dispersion relations of the optical and acoustic modes for forward and backward SBS are shown FigS.\ref{FigS2}a,b, respectively. 

In forward SBS, the Brillouin nonlinearity occurs between the co-propagating optical waves. The pump light (blue circle) is coupled to the Stokes light (red circle) and generates a FSBS acoustic phonon with a frequency of $f_{fa} = \nu_p - \nu_{fs}$ and a wave-vector of $q_f=k_p-k_{fs}$ (FigS.\ref{FigS2}a). Here, $\nu_{i}$, $k_{i}$ ($i$ = $p$, $fs$) are the frequencies and wave-vectors of the pump wave ($p$), and the forward Stokes wave ($fs$). Due to the significantly slower acoustic velocity in comparison to optical velocity, the wave-vector of the acoustic wave in FSBS, denoted as $q_f$, is extremely small (near zero). And the group velocity $v_g$, equivalent to the slope of the tangent line to the dispersion curve (gray dashed lines in FigS.\ref{FigS2}b), is generally slower for FSBS.

In backward SBS, the Brillouin nonlinearity occurs between the counter-propagating pump wave (blue circle) and Stokes wave (yellow circle), as illustrated in FigS.\ref{FigS2}a. The resulting phonon wave-vector is $q_b=k_p+k_{bs}$, almost twice as large as $k_p$. It is much larger than $q_f$. Therefore, to satisfy the phase-matching requirement of BSBS, the acoustic mode generally has a higher eigenfrequency, as shown in FigS.\ref{FigS2}b. Additionally, the backward acoustic mode exhibits a larger group velocity compared to FSBS. It implies that, with comparable phonon lifetimes, BSBS phonons exhibit significantly longer propagation distances compared to FSBS phonons.

\subsection{Brillouin gain model}
For pump power ($P_p$) and Stokes power ($P_s$), the coupled differential equations along the propagation direction (z-axis) with linear and nonlinear losses are [1]:
\begin{align}
\label{gainModel}
\frac{1}{P_p(z)}\frac{dP_p(z)}{dz}=-G_B(\Omega) P_s(z)-\alpha -\beta P_p(z)-\gamma P_p(z)^2,\\
\pm\frac{1}{P_s(z)}\frac{dP_s(z)}{dz}=G_B(\Omega) P_p(z)-\alpha-2\beta P_p(z)-\gamma P_p(z)^2,
\end{align}
where the $\pm$ corresponds to forward and backward SBS, respectively. $G_B(\Omega)$ is the Brillouin gain coefficient and can be expressed as [8]:
\begin{equation}
\label{GB}
G_B(\Omega)=\frac{2 \omega_p Q_m L(\Omega)}{m_{eff}\Omega^2_a}\lvert\int f_{ES}dA + \int f_{RP}dl \rvert^2,
\end{equation}
where $\omega_p$ is the angular frequency of the pump light. $Q_m$ and $\Omega_a$ are respectively the mechanical quality factor and the angular eigenfrequency of the acoustic mode. $L(\Omega)=(\Gamma/2)^2/[(\Gamma/2)^2+(\Omega-\Omega_a)^2]$ is the Lorentzian gain profile for an acoustic mode with the acoustic damping rate $\Gamma$. $m_{eff}=\int \rho\lvert \mathbf{u}\rvert^2/max\lvert \mathbf{u}\rvert^2dA$ is the effective mass of an acoustic mode with displacement field $\mathbf{u}$ and density $\rho$. The last two items are the area overlap integrals of the electrostrictive force $f_{ES}$ and line overlap integrals of the radiation pressure $f_{RP}$, respectively. Dividing $Q_m$ to the left side of Eq.\ref{GB} yields the coupling factor $G_B/Q_m$, which reflects the overlap integral of optical and acoustic modes. The simulations and optimizations of the coupling factor in Methods are based on this expression.

\begin{figure*}[ht]
\centering
\captionsetup{singlelinecheck=no, justification = RaggedRight}
\hspace*{-1.0cm}
\includegraphics[width=16cm]{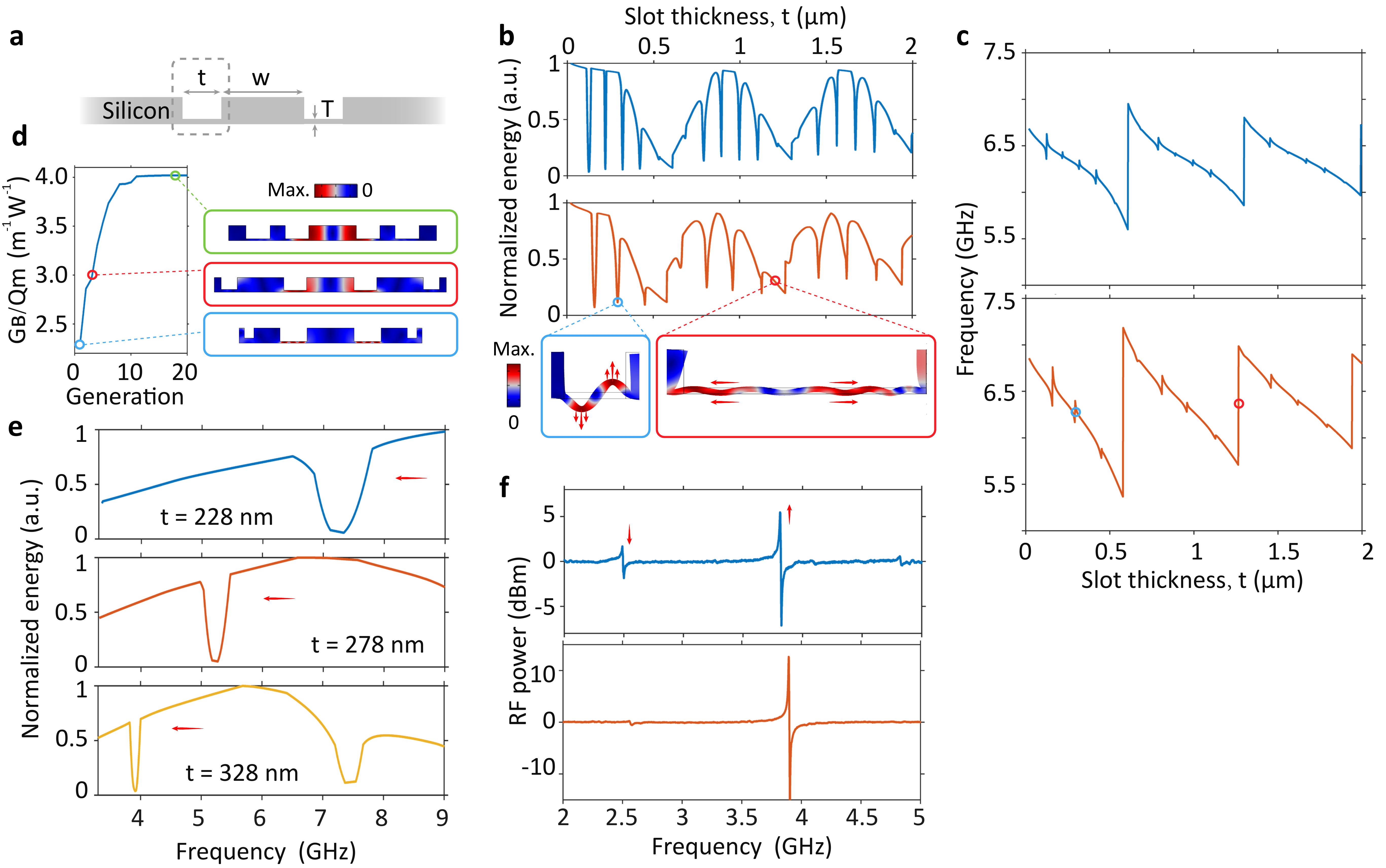}
\caption{\textbf{a} Schematic of etched slots on a silicon membrane. \textbf{b} The relationship between the slot thickness and the normalized acoustic mode energy in the waveguide. The deformation in the blue and red boxes is magnified proportionally according to the displacement field. \textbf{c} The relationship between the slot thickness and the eigenfrequency of the Brillouin-active mode. The blue and red circles on the orange line show the resonance condition for shear mode and longitudinal mode, corresponding to the acoustic modes marked by the circles and boxes on the orange line in figure \textbf{b}. In figure \textbf{b,c}, the blue and orange lines correspond to the remaining silicon layer thickness T = 15 and 30nm, respectively. \textbf{d} The optimization process of the genetic algorithm and the corresponding variation of the elastic displacement field. \textbf{e} Normalized acoustic energy with varied acoustic frequency. Three cases with slot thickness t = 228, 278, 328 nm are shown. \textbf{f} Illustration of frequency selectivity in SARAWs. The top is the heterodyne FWM measurement result before optimization; the bottom is the heterodyne FWM measurement result for SARAWs with enhanced phonon mode near 4 GHz.
}
\label{FigS3}
\end{figure*}

\section{Anti-resonance condition and optimization using genetic algorithm}
\subsection{Anti-resonance condition}
The acoustic modes in waveguides can be obtained by solving the elastic equation of motion [9]
\begin{equation}
\label{elasticMotion}
\rho\frac{d^2\mathbf{u}}{dt^2}=\nabla\cdot\mathbf{T},
\end{equation}
where $\rho$ is the density, $\mathbf{u}$ is the displacement field and $\mathbf{T}$ is the stress tensor. For an ideal planar waveguide (upper part of Fig.1a in the main text), the entire structure can be decomposed into layers with different acoustic velocities. Eq.\ref{elasticMotion} can be applied in each layer and solved numerically with appropriate boundary conditions. Analogous to an optical Fabry-Pérot cavity, under ideal conditions, the acoustic resonance and anti-resonance conditions can be approximately expressed as [9]: 
\begin{align}
\label{EQ1}
k_s t &= n\pi, n\in\mathbb{Z}^{+},\\
k_s t &= (n-0.5)\pi, n\in\mathbb{Z}^{+},
\end{align}
respectively, where $k_s$ is the transverse wave-vector. Acoustic resonance and anti-resonance states appear periodically and alternately, as slot thickness t increases. Under the anti-resonance states, the acoustic waves in the slow-velocity layers experience destructive interference, confining the acoustic field within the central layer. On the other hand, the acoustic waves are primarily distributed in the slow-velocity layers under the resonance states.

Due to complex boundary conditions and structural distortions of the etched slots, the resonance and anti-resonance conditions of SARAWs can not be directly solved using Eq.\ref{EQ1}. Therefore, we employ the finite element solver COMSOL to numerically solve the acoustic frequencies and acoustic mode distributions under various geometric structures.

In order to further illustrate the resonance and anti-resonance conditions of SARAWs and demonstrate the corresponding acoustic mode distributions, we set the waveguide width W to 700 nm and scanned the slot thickness t up to 2 $\mu$m (FigS.\ref{FigS3}a). To evaluate the acoustic energy confinement effect under different parameters, we tracked normalized energy, defined as the ratio of acoustic mode energy in the central waveguide to the total acoustic mode energy, as shown in FigS.\ref{FigS3}b. And we performed scans for two cases with the remaining thickness of the silicon layer T = 15 nm (blue lines) and 30 nm (orange lines). The bottom boxes of FigS.\ref{FigS3}b display two types of acoustic modes of the etched slots under resonance conditions, where the left mode (blue box) can be interpreted as a shear mode along the slot in the horizontal direction, and the right mode (red box) can be interpreted as a longitudinal mode along the slot in the horizontal direction. To realize acoustic anti-resonance in SARAWs, it has to meet the anti-resonance conditions of the shear and longitudinal modes simultaneously.

As slot thickness t increases (FigS.\ref{FigS3}b), the variation of the normalized energy is a superposition of two distinct periodic oscillation profiles. For the case of the remaining thickness of the silicon layer T = 15 nm, the narrower dips of the normalized energy curve correspond to the resonance conditions of the shear mode, with a repetition period of approximately 100 nm. Meanwhile, the wider dips correspond to the resonance conditions of the longitudinal mode, with a repetition period of approximately 700 nm. Compared to the case of T = 30 nm, the shear mode exhibits a notable relative change in the repetition period, increasing from 100 nm to 150 nm. On the other hand, the longitudinal mode experiences a slight shift in the repetition period, decreasing from 700 nm to 670 nm. This disparity is primarily attributed to the pronounced influence of the remaining silicon layer thickness T on the shear wave velocity. In contrast, the longitudinal wave velocity remains relatively constant, resulting in the stable resonance conditions of the longitudinal mode. Since the eigenfrequency of Brillouin-active acoustic mode in the central waveguide with waveguide width W = 700 nm is approximately 5.9 GHz, the velocities of the shear mode in the etched slots are 1755 and 1170 m/s, corresponding to T = 30 and 15 nm respectively. The fact, that the velocity in etched slots decreases as the T decreases, enables etched slots to form equivalent slow velocity layers for anti-resonant reflecting. It acts as the basic principle for constructing anti-resonant structures in SARAWs. The uneven appearance of resonance conditions in Fig.2b-d in the main text also originates from this. As SARAWs are fabricated using the loading-effect-based technique, wider slot thickness results in smaller T, causing the resonance and anti-resonance conditions to drift.
 
FigS.\ref{FigS3}c illustrates the frequency of the Brillouin-active acoustic mode as a function of the slot thickness t. When the resonance conditions for shear or longitudinal modes are met, the Brillouin-active acoustic mode in the central waveguide and the acoustic modes (shear or longitudinal) in the etched slots are strongly coupled. The phenomenon of avoided mode crossing is observed in both figures [10]. 

For structural stability, we have set the slot thickness (t) close to 250 nm, aligning with the first anti-resonance point of the longitudinal mode. The overall trend of the acoustic frequency and the coupling factor ($G_B/Q_m$) is only slightly affected by the longitudinal mode with increasing slot thickness (t) (Fig.2b-d in the main text). Consequently, we only focus on the anti-resonance condition of the shear mode in the main text.

\subsection{Optimization of SARAWs using genetic algorithms}
Considering the influence of the loading effect during the etching process, the theoretical prediction of anti-resonance conditions becomes complex. Additionally, obtaining the optimal structural parameters through parameter scans is time-consuming and impractical. Therefore, we employ the genetic algorithm to streamline the optimization process. We take the coupling factor ($G_B/Q_m$) as the fitness function for optimization. Because achieving the maximum coupling factor requires SARAWs to simultaneously confine both acoustic and optical modes, genetic algorithms will automatically optimize geometric parameters to satisfy the anti-resonance conditions of the acoustic mode.

As an example, FigS.\ref{FigS3}d illustrates the optimization process for SARAWs with $W_1$ = 700 nm. The genetic algorithm converges after 10 generations, and the acoustic displacement field is squeezed into the central waveguide (green box), corresponding to the optimal coupling factor. The optimized parameters ($W_1$, $W_2$, $W_3$, $t_1$, $t_2$, $t_3$, $d_1$, $d_2$, $d_3$ (unit nm)) corresponding to Fig.2h in the main text are (450, 138, 189, 242, 266, 1,500, 194, 196, 220), (700, 223, 255, 346, 356, 1,500, 193, 194, 220), and (1,200, 394, 402, 500, 419, 1,500, 200, 197, 220), respectively. The parameters of the SARAW with W$_1$ = 700 nm in Fig.1,2 in the main text are also derived from here.

We emphasize that the genetic algorithm offers a rapid approach to optimize targeted parameters. By designing the fitness function appropriately, we can optimize not only the coupling factor, but also the mechanical quality factor and mode field profile for specific acoustic modes based on the application scenario.
 
\subsection{Frequency selectivity of SARAWs}
To further illustrate the acoustic frequency selectivity of SARAWs, we swept the central waveguide width W (FigS.\ref{FigS3}a) to obtain varied eigenfrequencies of the Brillouin-active acoustic mode. By controlling W to scan the acoustic frequency from 4 to 9 GHz (FigS.\ref{FigS3}e), we tracked the variation of the normalized acoustic mode energy of the central waveguide for different slot thicknesses t with the remaining silicon layer thickness T of 30 nm. The blue, orange, and yellow curves represent the cases of slot thickness t = 228, 278, and 328 nm, respectively. Clear peaks and dips appear in the curves, corresponding to the anti-resonance and resonance frequencies with the given slot thickness. Anti-resonant peaks serve to enhance the acoustic energy, while resonant dips act as inhibitors, suppressing the acoustic energy. Leveraging this characteristic, we can selectively filter or suppress acoustic modes at specific frequencies. As the slot thickness (t) increases, both the resonance and anti-resonance frequencies shift to lower values, and a second resonance frequency emerges when t = 328 nm. This resembles the dependence of the free spectral range (FSR) on the length of the Fabry-Pérot cavity, and enables us to manipulate the frequency selectivity by adjusting the anti-resonant structures.

Utilizing this frequency selectivity, SARAWs can selectively enhance a specific acoustic mode while suppressing others. For a particular SARAW of $W_1$ = 1,200 nm, it can support multiple acoustic modes, resulting in the appearance of multiple peaks in the measurement result of the heterodyne FWM experiment (the top of FigS.\ref{FigS3}f). By adjusting the anti-resonant structures, the SARAW can suppress the 2.5 GHz mode and enhance the acoustic mode near 4 GHz (the bottom of FigS.\ref{FigS3}f). Leveraging the property of selectively enhancing a specific acoustic mode and suppressing others, SARAWs offer a more versatile design dimension.

\begin{figure*} [!ht]
\centering
\captionsetup{singlelinecheck=no, justification = RaggedRight}
\includegraphics[width=17cm]{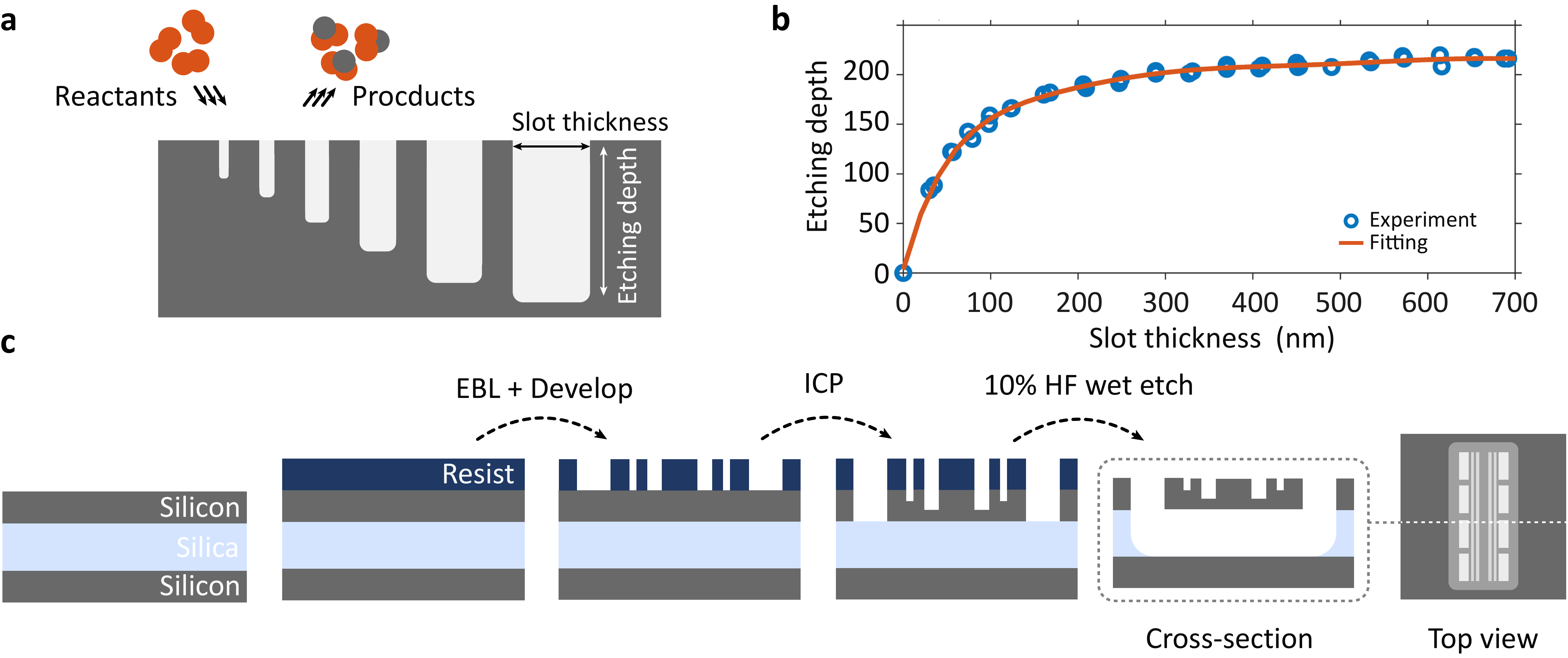}
\caption{\textbf{Inductively coupled plasma etching based on loading effect. a} Diagram of loading effect. \textbf{b} Experiment and fitting of loading effects. \textbf{c} Fabrication process.
}
\label{FigS4}
\end{figure*}

\begin{figure*} [ht]
\centering
\captionsetup{singlelinecheck=no, justification = RaggedRight}
\includegraphics[width=16cm]{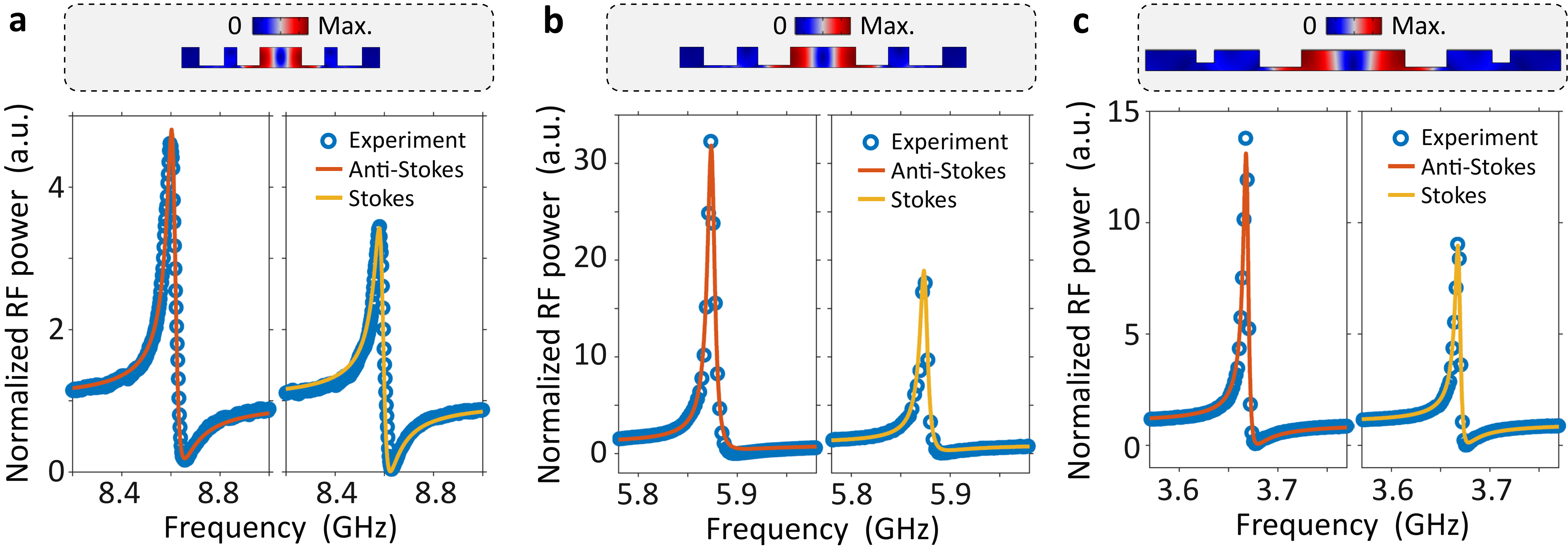}
\caption{\textbf{Simulation and experimental results of the heterodyne four-wave mixing experiment. a-c} The elastic displacement magnitude (upper) and the fittings of the experimental results for optimized SARAWs are depicted for waveguide widths $W_1$ = 450, 700, and 1,200 nm, respectively.
}
\label{FigS5}
\end{figure*}

\section{Inductively coupled plasma etching process based on loading effect}
Loading effect is an etching phenomenon in which the etching rate depends on the aperture size due to the consumption of reactants, as illustrated in FigS.\ref{FigS4}a. The etching rate increases with wider aperture width. This effect is also referred to as reactive ion etching (RIE) lag or aspect ratio dependent etching (ARDE) [11]. In this work, we fabricated a series of slots with varying thicknesses and measured the etching depths to obtain the loading effect curves shown in FigS.\ref{FigS4}b. The slot thickness in the anti-resonant structure (t$_1$ and t$_2$ in Fig.1c in the main text) is typically 300 $\sim$ 600 nm, resulting in the etching depth of 170 $\sim$ 200 nm, which prevents over-etching and ensures a stable connection for SARAWs. We emphasize that this curve (FigS.\ref{FigS4}b) can be adjusted by tuning the inductively coupled plasma (ICP) recipe parameters, such as reactant species and ICP power. The curve exhibited good reproducibility. We incorporated it into the parameter constraints in the genetic algorithm, and achieved excellent agreement between the fabricated and simulated parameters. 

Based on the loading effect, the fabrication process of SARAWs is depicted in FigS.\ref{FigS4}c. First, a positive electron beam resist is spin-coated, followed by electron beam lithography (EBL) to transfer the designed pattern. After development, the pattern is etched by ICP. Due to the loading effect, the wider slots at the edges (t$_3$ in Fig.1c in the main text) are overetched to expose the underlying silica, which is then removed by wet etching with 10$\%$ hydrofluoric acid solution, thus achieving full suspension of the SARAW. With this fabrication process, only one exposure and etching step is required, eliminating the need for overlay exposure. It leads to very high fabrication precision for SARAW. According to our fabrication results, a sub-5 nm feature pitch can be achieved.

\section{Heterodyne four-wave mixing experiment}
The spectral shape of the heterodyne four-wave mixing measurement (Fig.2a in the main text) arises from the interference of the resonant Brillouin scattering and the background Kerr nonlinearity. It follows a Fano-like line shape given by [1, 12]:
\begin{equation}
\label{FWM}
f_L(\Omega)=\left| e^{i\varphi}+\frac{G_B L_{SBS}}{4\gamma_k L_{tot}}\frac{\Omega_a/(2Q_m)}{\Omega_a-\Omega-i\Omega_a/(2Q_m)} \right|^2,
\end{equation}
where $\varphi$ is a relative phase between the Brillouin and Kerr nonlinearities, $L_{SBS}/L_{tot}$ is the ratio of Brillouin active length to total waveguide length, and $\gamma_k$ is the Kerr nonlinear coefficient. Based on the full vectorial model from Ref.[13], we obtain the $\gamma_k$ values of SARAWs with waveguide widths $W_1$ = 450, 700, and 1,200 nm in Fig.2h in the main text, which are 257, 179, and 110 $W^{-1}m^{-1}$, respectively.

Applying Eq.\ref{FWM} to fit the experimental results of heterodyne FWM measurement, we can derive the values of $G_B$ and $Q_m$. For the three waveguide widths in Fig.2h in the main text, the fitting results of the Stokes and anti-Stokes spectra are shown in FigS\ref{FigS5}a-c. The corresponding values 
of $G_B$ ($W^{-1}m^{-1}$) and $Q_m$ are (1,800, 210); (3,530, 680); (1,860, 850), for waveguide width $W_1$ = 450, 700, and 1,200 nm, respectively.

\begin{figure}[h]
\centering
\captionsetup{singlelinecheck=no, justification = RaggedRight}
\hspace*{-0.5cm}
\includegraphics[width=9.0cm]{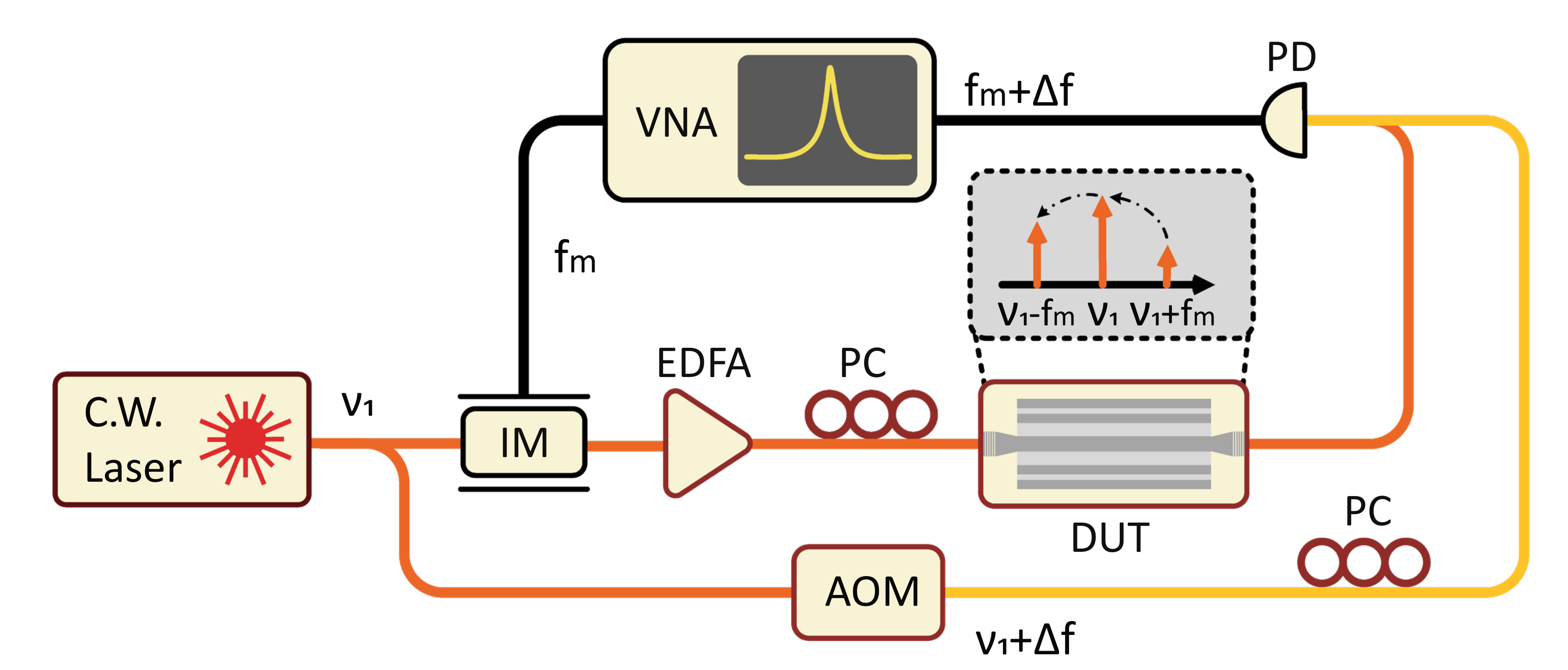}
\caption{\textbf{The setup of three-tone direct gain experiment.} Abbreviations: C.W. Laser, continuous-wave lasers; IM, intensity modulator; VNA, vector network analyzer; EDFA, erbium-doped fiber amplifier; DUT, device under test; AOM, acousto-optic modulator; PD, photodetector; PC, polorization controller.
}
\label{FigS6}
\end{figure}

\section{Three-tone direct gain experiment}
The three-tone gain experiment [12] has a simpler setup than the traditional small-signal gain experiment. The setup in this work is shown in FigS.\ref{FigS6}. The C.W. laser (frequency $\nu_1$, wavelength 1,550 nm) is divided into two paths via a coupler. The upper branch passes through an IM and generates two sidebands ($\nu_1 \pm f_m$). The carrier acts as the pump light, and the sidebands act as the probe lights. To satisfy the small-signal condition, the sideband powers are set to be 25 dB lower than the carrier. The light in the bottom path experiences a frequency shift to $\nu_1+\Delta f$ via an AOM, and acts as a reference signal. In the DUT, SBS facilitates energy transfer between the pump and the two sidebands when $f_m$ matches the Brillouin frequency shift. The strength of the energy transfer is directly related to $G_B$. After passing through the DUT, the light ($\nu_1\pm f_m$) is then combined with the reference signal ($\nu_1+\Delta f$) in the PD. The intensity variation of the beat signal $f_m+\Delta f$ is recorded by VNA, corresponding to the optical Stokes sideband $\nu_1-f_m$. 

In this process, the coupled amplitude equations for the pump ($a_p$), Stokes ($a_s$), anti-Stokes ($a_{as}$) and acoustic ($a_a$) waves can be expressed as [12]
\begin{align}
\label{Threetone}
\frac{d a_s}{dz}&=\frac{G_B\Gamma}{4}\chi^*a_pa_a^*-\frac{1}{2}(\alpha+2\beta\left|a_p\right|^2+\gamma\left|a_p\right|^4)a_s,\notag\\
\frac{d a_p}{dz}&=-\frac{G_B\Gamma}{4}(\chi a_sa_a-\chi^*a_{as}a_a^*)\notag\\
&\hspace{0.5cm}-\frac{1}{2}(\alpha+\beta\left|a_p\right|^2+\gamma\left|a_p\right|^4)a_p,\notag\\
\frac{d a_{as}}{dz}&=-\frac{G_B\Gamma}{4}\chi a_pa_a-\frac{1}{2}(\alpha+2\beta\left|a_p\right|^2+\gamma\left|a_p\right|^4)a_{as},\notag\\
\hspace{0cm}a_a&=a_s^*a_p+a_p^*a_{as}.
\end{align}
Here, the frequency response $\chi=1/(\Gamma/2+i(\Omega_a-\Omega))$. The optical powers $P_i=\left|a_i\right|^2$ (i = s, p, as).

In traditional small-signal gain experiment, only a single Stokes light is involved, and the small-signal gain is defined as the amplification of the probe light. In contrast, the three-tone gain experiment utilizes the coherent excitation of the acoustic field by both sidebands ($\nu_1\pm f_m$) [12]. This leads to a higher Stokes gain for the sideband ($\nu_1-f_m$) compared to the small-signal gain. Therefore, the three-tone experiment is more effective for detecting the smaller Stokes gain. However, to derive the corresponding traditional small-signal gain, we need to numerically solve Eq.\ref{Threetone} and fit the experimental data from three-tone experiment (Fig.3b in the main text). The obtained parameters are used to determine the small-signal gain using Eq.S5-6 (Fig.3d in the main text). Moreover, it should be noted that the mechanical quality factor $Q_m$ can not be directly determined by the full width at half maximum (FWHM) of Fig.3a in the main text. Instead, it should be derived by numerically solving Eq.\ref{Threetone} and fitting the experimental results.

\section{The backward SBS experiment}
\begin{figure}[h]
\centering
\captionsetup{singlelinecheck=no, justification = RaggedRight}
\hspace*{-0.5cm}
\includegraphics[width=9.0cm]{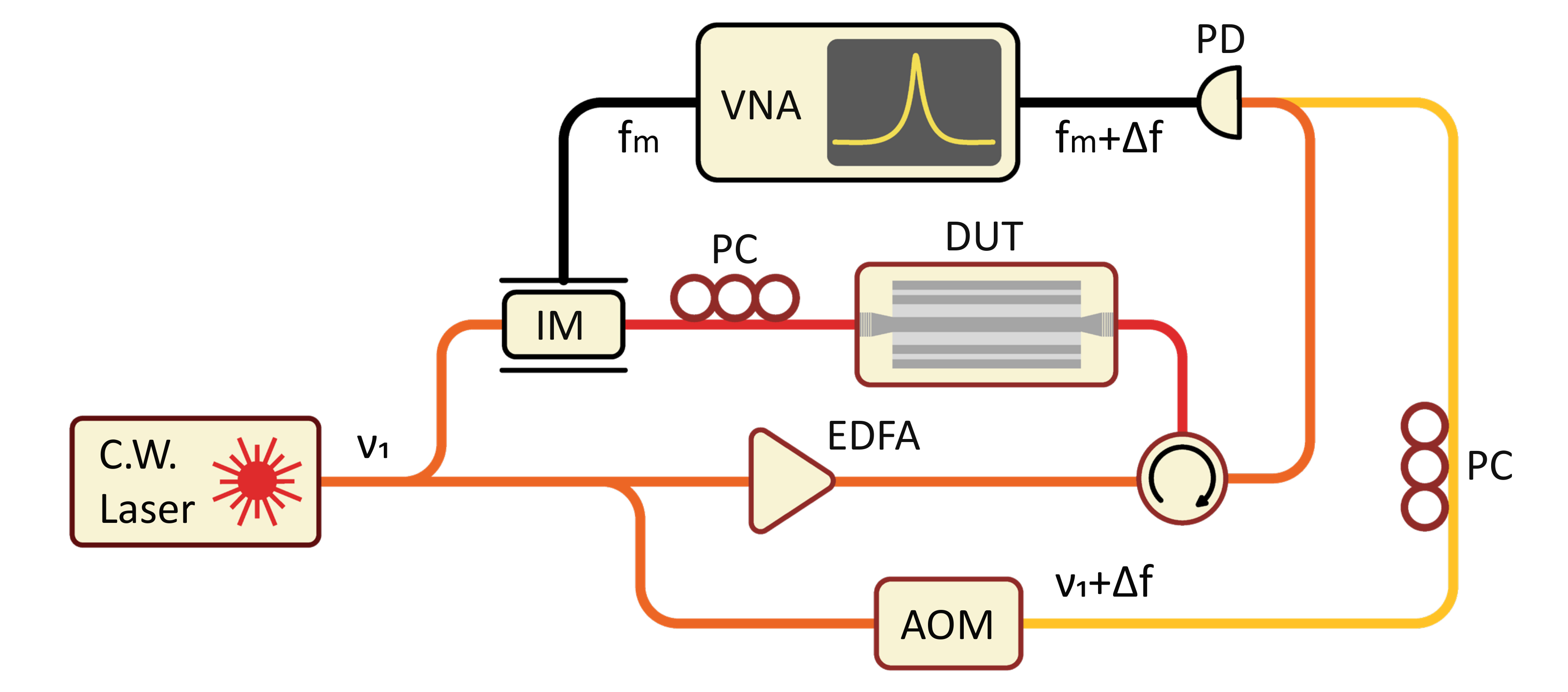}
\caption{\textbf{The setup of the backward SBS experiment.}
}
\label{FigS7}
\end{figure}

The setup of the backward SBS experiment is shown in FigS.\ref{FigS7}. The light from a C.W. laser (frequency $\nu_1$, wavelength 1,550 nm) is split into three paths. The light in the upper path passes through an IM and generates a sideband at $\nu_1-f_m$ (25 dB below the carrier to meet small-signal condition), serving as the probe light. The light in the middle path is amplified by an EDFA and acts as the pump light, entering the DUT from the right through a circulator. When $f_m$ matches the backward Brillouin frequency shift, the probe light undergoes amplification through the SBS process. After passing through the DUT, the probe light ($\nu_1-f_m$) is further mixed with a reference signal at $\nu_1+\Delta f$ from an AOM in the lower path branch. The beat signal ($f_m+\Delta f$) is measured by the VNA, and its intensity corresponds to the small-signal gain of backward SBS.

\begin{figure*} [p]
\centering
\captionsetup{singlelinecheck=no, justification = RaggedRight}
\includegraphics[width=16cm]{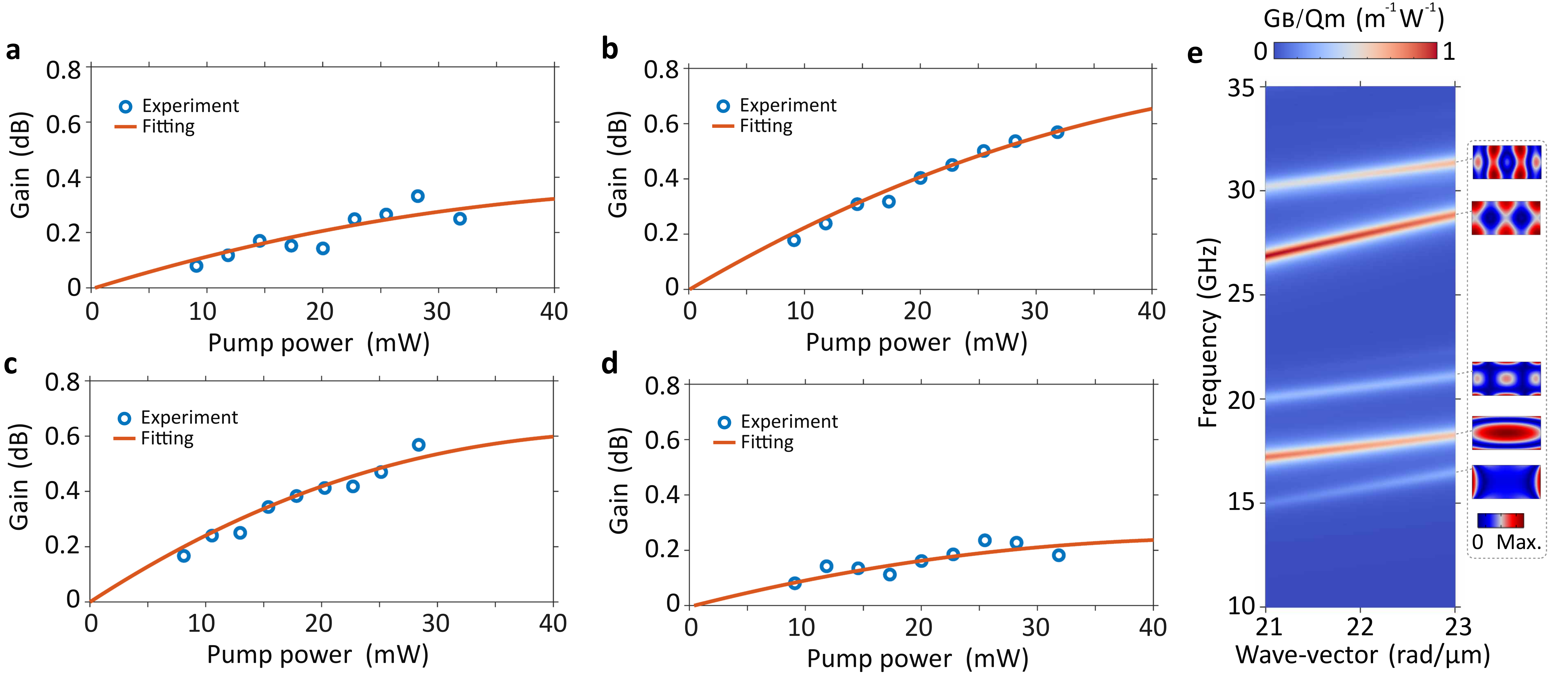}
\caption{\textbf{Fitting of backward Brillouin gain coefficient and acoustic dispersion. a-d} Backward Brillouin gain coefficient fitting of SARAWs with different parameters, corresponding to the acoustic modes shown in Fig.4a-d in the main text. \textbf{e} The dispersion relation of the backward SBS acoustic modes. The displacement magnitude of the acoustic modes is shown in the gray dashed boxes.
}
\label{BSBS}
\end{figure*}

\begin{figure*} [h]
\centering
\captionsetup{singlelinecheck=no, justification = RaggedRight}
\includegraphics[width=16cm]{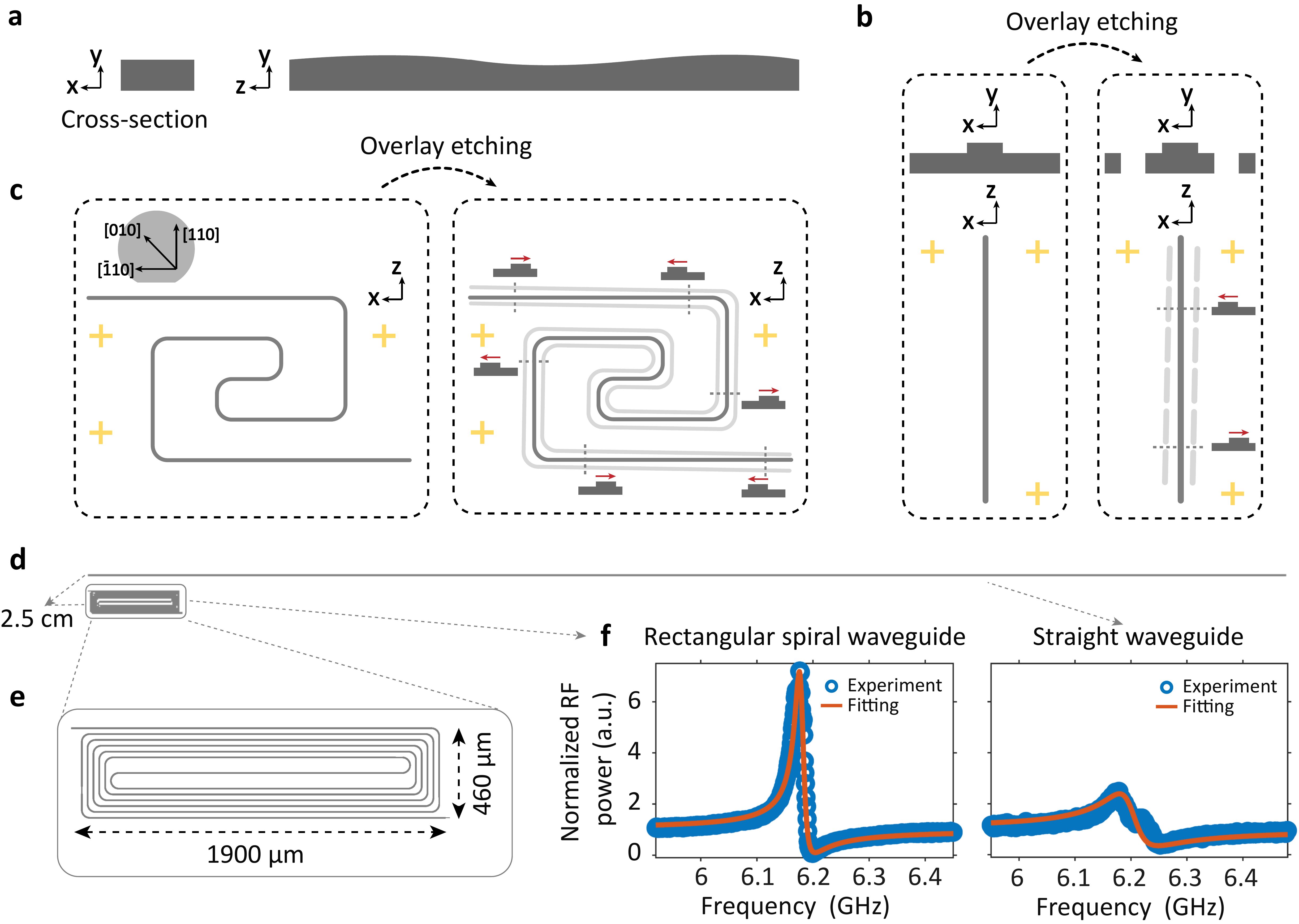}
\caption{\textbf{Inhomogeneous broadening and peak splitting. a} The non-uniformity of silicon membrane. \textbf{b,c} The mismatches of straight and spiral waveguides induced by overlay exposure and etching. The golden crosses indicate the marks required for alignment, the dark grey lines denote rib waveguides, and the light grey lines represent the overlay etched regions. The upper left corner of the dashed box on the left side of figure \textbf{c} indicates the SOI wafer and the crystal orientations. \textbf{d,e} Schematics of the footprints of straight and rectangular spiral waveguides, and the enlarged view (\textbf{e}). \textbf{f} The heterodyne FWM experiment results of straight (right) and rectangular spiral (left) waveguides.
}
\label{Broaden}
\end{figure*}

To obtain the $G_B$ of backward SBS, we swept the pump power and measured the corresponding small-signal gain. Solving the Eq.S6 numerically and fitting the experimental data, the obtained $G_B$ (FigS.\ref{BSBS}a-d) are 530, 600, 430, and 480 $W^{-1}m^{-1}$, corresponding to Fig.4a-d in the main text respectively.

According to the phase-matching relationship (FigS.\ref{FigS2}a), the group velocity of the acoustic mode, can be inferred by the slope of the tangent line to the acoustic dispersion curve. The acoustic modes depicted in FigS.\ref{BSBS}a-d are similar to those in a suspended rectangular waveguide. Therefore, for the sake of simulation convenience, we simulate the acoustic dispersion of the suspended rectangular waveguide (220 nm $\times$ 450 nm) near a wave-vector of $2k_p$, as shown in FigS.\ref{BSBS}e. The 18 and 28 GHz acoustic modes are consistent with the backward SBS acoustic modes observed in SARAWs. Their group velocities are 3,190 and 6,980 m/s, respectively. With a $Q_m$ of 1,960 in Fig.4d in the main text, we estimate the acoustic decay length of the bell-shaped mode to be larger than 50 $\mu m$. The acoustic mode at the bottom of the gray dashed box in FigS.\ref{BSBS} corresponds to the Fabry-Pérot acoustic mode of backward SBS in nano wire on pillar [14]. Owing to its small $G_B/Q_m$, this mode is not the focal point of our investigation.

\section{Inhomogeneous broadening and peak splitting}
The inhomogeneous broadening of Brillouin resonance caused by fabrication defects can be classified according to the spatial scale of the defects [15]. Defects with short-scale variations (compared with the acoustic decay length), such as sidewall roughness, may result in large acoustic energy dissipation and decline in the mechanical quality factor. On the other hand, defects with long-scale variations, such as non-uniformity of silicon membrane thickness (FigS.\ref{Broaden}a), can induce drifts in the geometric parameters of the waveguide cross-section, thereby causing shifts in the acoustic frequencies. In extreme cases, it will result in peak splitting of Brillouin resonance, inducing a dramatic decrease in Brillouin gain. 

Short-scale defects can be mitigated by improving the EBL resist and etching process, as well as thermal oxidation. For long-scale defects, further analysis is performed in this work. Since the straight waveguide layout (FigS.\ref{Broaden}a) spans a longer distance, it is more susceptible to long-range fluctuations. The spiral waveguide with a smaller footprint is expected to show less inhomogeneous broadening. However, as reported in previous work [12], even more severe inhomogeneous broadening is observed in spiral waveguides compared to straight waveguides with the same length. 

It implies that other factors induce inhomogeneous broadening in the spiral waveguide and take on a dominant role. The reasons for this phenomenon can be described as follows. For waveguide structures that require overlay exposure, such as suspended rib waveguides [1], alignment of the marks in overlay exposure introduces mismatches between the pattern layers (FigS.\ref{Broaden}b). On our fabrication platform, the mismatches are larger than 20 nm, which causes drift in the waveguide cross-sectional geometry and results in inhomogeneous broadening. In particular, for spiral waveguides, this mismatch is dispersed throughout the spiral, leading to more severe inhomogeneous broadening and even peak splitting (FigS.\ref{Broaden}c). However, SARAWs fabricated based on the loading-effect etching technique can avoid overlay exposure, thus eliminating the inhomogeneous broadening from alignment mismatches.

Moreover, the inhomogeneous broadening in spiral waveguides can also arise from the variation in crystal orientation. The eigenfrequencies of acoustic modes shift in different crystal orientations. In the case of spiral waveguides, such as Archimedean spirals, the waveguide direction shifts between [110] and [010] orientations. This variation in orientation leads to noticeable inhomogeneous broadening and peak splitting. To mitigate inhomogeneous broadening induced by crystal orientation, we opt for a rectangular spiral layout in SARAWs, which have the same crystal orientation in the orthogonal directions.

To prove the effectiveness of the method mentioned above, we fabricated 2.5-cm-long SARAWs with $W_1$ = 700 nm. We compare forward SBS results of the heterodyne FWM experiment measured in the rectangular spiral ($G_B$ = 1,690 $W^{-1}m^{-1}$, $Q_m$ = 325) and straight ($G_B$ = 678 $W^{-1}m^{-1}$, $Q_m$ = 90) SARAWs (FigS.\ref{Broaden}f). Compared to straight SARAW, spiral SARAW exhibits not only a smaller footprint, but also an enhancement of 2.5 times in the gain coefficient and a 3.6 times increase in the mechanical quality factor. 

\newpage
\noindent[1] Kittlaus, E. A., Shin, H. \& Rakich, P. T. Large Brillouin amplification in silicon. Nature Photonics 10, 463–467 (2016).

\noindent[2] Gao, M. et al. Probing material absorption and optical nonlinearity of integrated photonic materials. Nature Communications 13, 3323 (2022).

\noindent[3] Wilson, D. J. et al. Integrated gallium phosphide nonlinear photonics. Nature Photonics 14, 57–62 (2020).

\noindent[4] Zhang, X. et al. Characterizing microring resonators using optical frequency domain reflectometry. Optics letters 46, 2400–2403 (2021).

\noindent[5] Payne, F. P. \& Lacey, J. P. A theoretical analysis of scattering loss from planar optical waveguides. Optical and Quantum Electronics 26, 977–986 (1994).

\noindent[6] Afshar, S. \& Monro, T. M. A full vectorial model for pulse propagation in emerging waveguides with subwavelength structures part i: Kerr nonlinearity. Optics express 17, 2298–2318 (2009).

\noindent[7] Hon, N. K., Soref, R. \& Jalali, B. The third-order nonlinear optical coefficients of si, ge, and si1- xgex in the midwave and longwave infrared. Journal of Applied Physics 110 (2011).

\noindent[8] Wiederhecker, G. S., Dainese, P. \& Mayer Alegre, T. P. Brillouin optomechanics in nanophotonic structures. APL photonics 4 (2019).

\noindent[9] Schmidt, M. K., O’Brien, M. C., Steel, M. J. \& Poulton, C. G. ARRAW: anti-resonant reflecting acoustic waveguides. New Journal of Physics 22, 053011 (2020).

\noindent[10] Stern, I., Carosi, G., Sullivan, N. \& Tanner, D. Avoided mode crossings in cylindrical microwave cavities. Physical Review Applied 12, 044016 (2019).

\noindent[11] Gosalvez, M. A. et al. Simulation of microloading and arde in drie. In 2015 Transducers-2015 18th International Conference on Solid-State Sensors, Actuators and Microsystems
(TRANSDUCERS), 1255–1258 (IEEE, 2015).

\noindent[12] Wang, K. et al. Demonstration of Forward Brillouin Gain in a Hybrid Photonic–Phononic Silicon Waveguide. ACS Photonics 8, 2755–2763 (2021).

\noindent[13] Afshar, S. \& Monro, T. M. A full vectorial model for pulse propagation in emerging waveguides with subwavelength structures part i: Kerr nonlinearity. Optics express 17, 2298–2318 (2009).

\noindent[14] Van Laer, R., Kuyken, B., Van Thourhout, D. \& Baets, R. Interaction between light and highly confined hypersound in 10 a silicon photonic nanowire. Nature Photonics 9, 199–203 (2015).

\noindent[15] Zurita, R. O., Wiederhecker, G. S. \& Alegre, T. P. M. Designing of strongly confined short-wave Brillouin phonons in silicon waveguide periodic lattices. Optics Express 29, 1736–1748 (2021).

\noindent
